\documentclass{jfm}

\usepackage{amsmath,amssymb}
\usepackage{mathrsfs}
\usepackage{xcolor}
\usepackage{bbm}
\usepackage{bm}
\usepackage{pict2e} % for making circle around number
\usepackage{booktabs}
\usepackage{tabularx}
\usepackage{setspace}
\usepackage{mathtools}
\usepackage{tikz}
\usepackage{soul}
\usepackage{graphicx,epstopdf}
\usepackage[colorlinks=true, linkcolor=blue, urlcolor=blue, citecolor=blue]{hyperref}
\usepackage{natbib}
\usepackage{newtxtext}
\usepackage{newtxmath}
\usepackage{natbib}
\usepackage{hyperref}
\usepackage{subcaption}
\usepackage{caption}
\hypersetup{
    colorlinks = true,
    urlcolor   = blue,
    citecolor  = black,
}

\newcommand{\RomanNumeralCaps}[1]

\begin{document}
%\newtheorem{lemma}{Lemma}
%\newtheorem{corollary}{Corollary}
%
%A method of fundamental solutions for micro/nano scale phase transitions: application of the coupled constitutive
%relations
\title{A second order constitutive theory for polyatomic gases: theory and applications}
\shorttitle{A second order constitutive theory for polyatomic gases}
\shortauthor{A. Rana,  S. Barve} %for header on even pages
\author
 {
 {Anirudh S. Rana\aff{1}
\corresp{
 \email{anirudh.rana@pilani.bits-pilani.ac.in}
},
}
 %Anil Kumar\aff{1}
\and
 Sukratu Barve\aff{2}
 }

\affiliation
{
\aff{1}
Department of Mathematics, BITS Pilani, Pilani Campus, Rajasthan 333031, India
\aff{2}
Department of Scientific Computing, Modeling and Simulation, Savitribai Phule Pune University, Pune 411007, India
}

\maketitle

\begin{abstract}
In the classical irreversible thermodynamics (CIT) framework, the Navier-Stokes-Fourier (NSF) constitutive equations are obtained so as they satisfy the entropy inequality, by and large assuming that the entropy flux is equal to the heat flux over the temperature. This article is focused on the derivation of second-order constitutive equations for polyatomic gases; it takes the basis of CIT, but most importantly, allowing up to quadratic nonlinearities in the entropy flux. Mathematical similarities between the proposed model and the classic Stokes-Laplace equations are exploited so as to construct analytic/semi-analytic solutions for the slow rarefied gas flow over different shapes. A set of second-order boundary conditions are formulated such that the model’s prediction for the drag force is in excellent agreement with the experimental data over the whole range of Knudsen number. We have also computed the normal shock structure in nitrogen for Mach $\mathrm{Ma} \lesssim 4$. A very good agreement was observed with the kinetic theory, as well as with the experimental data.
\end{abstract}

\section{Introduction}
\label{sec:intro}

%   Polyatomic gases are often encountered in nanoscale mechanical devices and systems. For example, few recent U.S. patents (see e.g., \citet{US009835363B2}, \citet{US20160295742A1}) have documented fabrication of nano-porous membranes that act as a carrier for an evaporating thin-film of a polyatomic fluid (for example, pentane in \citep{US009835363B2}), allowing for very efficient thermal management systems in ways that are not feasible with traditional cooling technologies. 
%   Besides, the gas most commonly encountered in everyday life is air. It is a mixture of gases---primarily nitrogen, which constitutes about 78\% of air, and oxygen at about 21\% along with trace amount of constituents, such as argon, carbon dioxide, ozone, and water vapor; most of these are polyatomic gases. 
  
 The gas most commonly encountered in everyday life is air. It is a mixture of polyatomic gases---primarily nitrogen, which constitutes about 78\% of air, and oxygen at about 21\% along with trace amount of constituents, such as argon, carbon dioxide, ozone, and water vapor; most of these are polyatomic gases. Air also contains tiny particles called aerosols, such as fog, dust, smoke and particulate air pollutants, so called respirable ultrafine particulates (UFPs). Owing to their nanoscale size, UFPs can penetrate deep into the lungs, blood streams and can have severe health implications including heart attacks and respiratory diseases \citep{COPD2017}. Aerosols can also be produced by coughing or sneezing and airborne viruses can pass from person to person through these nanoscale droplets; therefore, scientists have paid much attention to dynamics of aerosols that waft through the air.   
 
 Modeling of micro/nano gas flows is especially intriguing because classical gas dynamics models fail to capture the flow characteristics (such as, drag force on aerosols, mass and heat flux on the evaporation meniscus) in rarefied conditions.
 As the representative physical length scale ($L$) of the flow becomes comparable to the mean free path ($\lambda$) in the gas, i.e., the Knudsen number Kn (= $\lambda/L$) $\approx 1$, Navier-Stokes-Fourier (NSF) equations fail to predict thermal and fluid flow fields to sufficient accuracy. 
%  The NSF equations are only valid when the Knudsen number is sufficiently small.
%  The Navier-Stokes-Fourier (NSF) equations with appropriate slip/jump boundary conditions may provide   are only valid when Kn $\ll 1$. 
 The Boltzmann equation \citep{Kremer2010} in such a situation offers a more accurate description of the micro/nano gas flows, but unfortunately is computationally expensive.
 As a work around over the last few years, an impressive body of research has been devoted to the development of extended hydrodynamic models for polyatomic gases \citep{zhdanov1968kinetic, Mallinger1998, Kremer2010, CAI201463, RahimiStruchtrup2014}, which may provide accurate and computationally efficient simulation-for-design capabilities down to nanoscales.

%   According to the value of Kn, the flow can be classified into four flow regimes, the continuum flow (Kn ≤ 0.001), the slip flow (0.001 < Kn ≤ 0.1), the transition flow (0.1 < Kn ≤ 10), and free molecular flow (Kn > 10).
% These models  that is not accessible by the conventional fluid dynamics.
% % provide a computationally cheaper and  not accessible by the conventional fluid dynamics.
 The notion of extended hydrodynamic models goes back to Grad's pioneering work on
approximating the Boltzmann equation via moment equations \citep{Grad1958}. 
% These
% equations are derived from the Boltzmann equation by multiplying it with
% polynomials in microscopic velocity and subsequently integrating it over the
% microscopic velocity space. 
%
The 13 moment equations
%%%%%%%%
%(five conservation laws and the
%balance equations for five %components of the deviatoric stress tensor and
%three components of heat flux vector) %%%%%
obtained via Grad's method yield unphysical sub-shocks and also prone to loose their hyperbolicity in strong non-equilibrium conditions. Struchtrup \& Torrilhon \citeyearpar{StruchtrupTorrilhon2003,TorrilhonStruchtrup2004} developed a linearly
stable system of regularized 13 moment (R13) equations for monoatomic gases
using the order-of-magnitude method \citep{Struchtrup2005MMS} by combining the ideas of the
Chapman--Enskog \citep{CC1970} and Grad's methods. The regularization process extends the validity of the Grad
closures through the addition of higher-order derivatives and provides smooth shock structures. \citet{RahimiStruchtrup2014} using the same
method obtained a hierarchy of moment equations for polyatomic gases at
different levels of accuracy. 
% At the first order the NSF equations were
% recovered whilst at the second order Grad 14 equations were obtained, which
% along with the classical 13 moment equations (five conservation laws and the
% balance equations for five components of the deviatoric stress tensor and
% three components of heat flux vector) also contained one balance equation
% for the dynamic temperature. At the third order accuracy, the regularized 19
% moment (R19) equations were obtained.
\citet{CAI201463} proposed a numerical regularized moment method of arbitrary order (NRxx) for polyatomic gases with ES-BGK collision operator.

Unfortunately, the extended hydrodynamic equations obtained by Grad's method, the Chapman--Enskog method  or the
order-of-magnitude method are not accompanied by a proper entropy balance
with strictly non-negative production. This might lead to breakdown of
solutions or unphysical results  \citep{StruchtrupNedler2020}.
Another approach towards approximation of the Boltzmann equation is rational
extended thermodynamics (RET) framework \citep{muller2013rational, ruggeri2015rational}, which, however, has the same limitation
as those for the moment equations (only the linearized equations have a proper
entropy law) Another popular family of moment closure methods is the
maximum-entropy hierarchy \citep{junk2002maximum, torrilhon2016modeling}. The maximum-entropy closures ensure entropy inequality and hyperbolicity but they suffer from high computational cost incurred to compute the closure. Furthermore, it is impossible to obtain a closed form solution from the higher order maximum-entropy closures.

% We re-consider the CIT framework which is based on the hypothesis of local equilibrium so that the Gibbs relation is assumed to be locally valid \citep{deGrootMazurBook1962}. The resulting transport equations are accompanied by a proper entropy inequality 
% % with non-negative entropy
% % generation 
% so that a proper thermodynamic structure is maintained. 
Recently,
a phenomenological procedure for monoatomic gases was proposed \citep{RGS2018} in which the entropy flux contains additional nonlinear contributions in stress and heat
flux---an ansatz consistent with the RET. The resulting coupled constitutive relations (CCR) are accompanied by an entropy inequality, in which the entropy remains the equilibrium entropy as integrated from the equilibrium Gibbs equation, but entropy flux and entropy generation exhibit higher order correction terms.
These additional terms in the entropy flux add a few second order correction terms
to the NSF system. 
The CCR system allows to formulate thermodynamically consistent boundary conditions and their
efficacy in capturing rarefaction effects  at small scales was shown by \citet{RGS2018, RSCLS2021}. 
% This success of the CCR theory, though limited to the case of monoatomic gases, is noteworthy \citep{RGS2018, RSCLS2021}.  
% Even the definitions of local thermodynamic quantities, especially temperature, play a critical role in generalization of the thermodynamic laws and relations to non-equilibrium scenarios. The assumption of internal energy proportional to local temperature is crucial in CIT. Thus local temperature ought to be defined unambiguously.  Entropy per unit mass  may also be expected to deviate from local equilibrium values. 
In this work, we extend the theory of the coupled constitutive relations to polyatomic gases.  The resulting model (polyatomic-CCR) is benchmarked  with experimental data for the slow rarefied gas flow over different shapes; the model's prediction for the drag force is in excellent agreement with  the experiments. The normal shocks structure in nitrogen is also computed for the Mach number $\mathrm{Ma} \lesssim 4$; again, there is a good match between the polyatomic-CCR model and the experiments.   
In large measures, the polyatomic-CCR model enables the study of small-scale physics that is not accessible by the conventional fluid dynamics at a very little to no added computational cost. 
% The CCR idea of modifying the entropy current constitutive law can still be fruitfully applied in the polyatomic case despite the limitations of applicability of local thermodynamic equilibrium.

The remainder of this paper is organized as follows. In Section 2, we introduce the
governing equations and CIT framework. In Section 2.3, derivation of the polyatomic-CCR model is presented. In Section 3, we introduce
linearized and dimensionless equations followed by their solutions for slow rarefied gas flow over a spherical and doublet particle in Section 4 and 6, respectively. A brief derivation of the Green's functions is also given in Section 5. In Section 6, normal stock structure in nitrogen gas is computed. We conclude and discuss future directions in Section 7.
\section{The governing equations}
\label{sec: the governing equations}
\subsection{Conservation laws for polyatomic gases}

The conservation laws are the balance equations for the mass density $\rho $, the fluid velocity 
 $\mathbf{v}$, internal energy $u$. For a compressible, polyatomic gas these are formulated as \citep{Kremer2010}
\begin{subequations}
\label{Conservation laws}
\begin{eqnarray}
\frac{\partial \rho }{\partial t}+\bigtriangledown \cdot \left( \rho \mathbf{%
v}\right)  &=&0\text{,}  \label{mass balance} \\
\frac{\partial \left( \rho \mathbf{v}\right) }{\partial t}+\bigtriangledown
\cdot \left[ \rho \mathbf{v\otimes v+}\left( p+\Pi \right) \mathbf{I+\sigma }%
\right]  &=&0\text{,}  \label{momentum balance} \\
\frac{\partial \left( \rho u+\frac{1}{2}\rho v^{2}\right) }{\partial t}%
+\bigtriangledown \cdot \left[ \left( \rho u+\frac{1}{2}\rho v^{2}+p+\Pi
\right) \mathbf{v+\sigma \cdot v+q}\right]  &=&0\text{,}
\label{energy balance}
\end{eqnarray}%
\end{subequations}
where $p$ is the pressure, $\Pi $ is the dynamic pressure, $\mathbf{\sigma }$ is the symmetric
and trace free part of the pressure tensor and $\mathbf{q}$ is the heat
flux. \ The internal energy $u$, of a polyatomic ideal gas is related to the
thermodynamics temperature $T$ through the Caloric equation $du=\frac{3+\delta }{2}d\theta:=\frac{d\theta}{\gamma -1}$ , i.e.,  the constant volume  specific heat is given by $c_v= \tfrac{3+\delta }{2}\mathrm{R}$. Here, $\mathrm{R}$ is the ideal gas constant and $\gamma$ $(= c_p/c_v)$ is the specific heat ratio.

% \begin{equation}
% du=\frac{3+\delta }{2}\mathrm{R}dT:=\frac{\mathrm{R}dT}{\gamma -1}
% \label{Caloric equation}
% \end{equation}%
For convenience, we also
define the temperature and the dynamic temperature in energy units as $\theta :=%
\mathrm{R}T\mathrm{\ }$and $\vartheta :=\Pi /\rho $.  
A value $\delta$  can be chosen such that the proper caloric equation of gas is recovered, i.e.,
\begin{equation}
    \delta = \frac{5-3 \gamma}{\gamma-1}.
\end{equation}
For example, for the nitrogen gas or for the dry air $\gamma \approx 1.4$, which gives $\delta=2$.

In molecular gases, $c_v$ depends on the temperature, which implies $\delta$ should also be a function of temperature. However, for processes where changes in the temperature are small, one can assume $c_v$ to be constant, hence, $\delta$ can be assumed constant. Throughout, this article we consider such processes, and assume $\delta$ to be constant.  Furthermore, the
pressure $p$ in (\ref{Conservation laws}) is assumed to given by the ideal gas law, $%
p=\rho \theta $. 

The conservation laws (\ref{Conservation laws}) need to be
closed by constitutive relations between the field variables ($\rho $, $%
\mathbf{v}$ and $\theta $) and their fluxes ($\Pi $, $\mathbf{\sigma }$ and $%
\mathbf{q}$). These constitutive relations can be obtained by enforcing the
second law inequality, as shown below.

\subsection{NSF constitutive relations from the CIT framework}

The CIT framework is based on the
hypothesis of local equilibrium, according to which, the Gibbs relation is
assumed to be locally valid, i.e., 
\begin{equation}
\theta d\eta =du-\frac{p}{\rho ^{2}}d\rho  \label{Gibbs relation}
\end{equation}%
where $\eta =s/\mathrm{R}$ denotes the dimensionless entropy ($s$ being the dimensional entropy). Multiplying the
last equation with $\rho $, and replacing $D\rho /Dt$, and $Du/Dt$ by means
of the conservation laws (\ref{Conservation laws}), one obtains the entropy
balance equation as 
\begin{equation}
\frac{\partial \left( \rho \eta \right)} {\partial t}+\bigtriangledown \cdot \left[
\rho \eta \mathbf{v+}\frac{\mathbf{q}}{\theta }\right] =-\frac{1}{\theta }%
\Pi \bigtriangledown \cdot \mathbf{v}-\frac{1}{\theta }\mathbf{\sigma :}%
\left\langle \bigtriangledown \mathbf{v}\right\rangle -\frac{\mathbf{q}}{%
\theta ^{2}}\mathbf{\cdot }\bigtriangledown \theta \text{.}
\label{entropy balance NSF}
\end{equation}%
Here, the angular parentheses around a matrix denotes the symmetric and
trace-free part, i.e., $\left\langle A\right\rangle =(A+A^{T})/2-\mathbf{I}%
Tr\left( A\right) /3$; $\mathbf{I}$ being the identity matrix and $Tr\left(
A\right) $ is the trace of $A$.

Comparing (\ref{entropy balance NSF}) with a general balance equation of the
form 
\begin{equation}
\partial \left( \rho \eta \right) /\partial t+\bigtriangledown \cdot \left(
\rho \eta \mathbf{v+\Psi }\right) =\Sigma \text{,}
\label{General balance equation}
\end{equation}%
it follows that the non-convective entropy flux $\mathbf{\Psi }=\mathbf{q}/\theta $ and the entropy
production
\begin{equation}
\Sigma =-\frac{1}{\theta }\Pi \underline{\bigtriangledown \cdot \mathbf{v}}-%
\frac{1}{\theta }\mathbf{\sigma :}\underline{\left\langle \bigtriangledown 
\mathbf{v}\right\rangle }-\frac{\mathbf{q}}{\theta ^{2}}\mathbf{\cdot }%
\underline{\bigtriangledown \theta }\text{.}
\label{Classic entropy production}
\end{equation}%
The expression for $\Sigma $ forms a bi-linear relationship in the
thermodynamic fluxes ($\Pi $, $\mathbf{\sigma }$, $\mathbf{q}$) and the
underlined thermodynamic forces. To guarantee the positiveness of the
entropy production $\Sigma $, it is sufficient to assume linear flux-force
relations of the form%
\begin{equation}
\mathbf{\sigma }%
=-2\mu \left\langle \bigtriangledown \mathbf{v}\right\rangle 
\text{, }\Pi =-\mu _{b}\bigtriangledown \cdot \mathbf{v}\text{\quad and \quad}%
\mathbf{q}=-\kappa \bigtriangledown \theta \text{,}  \label{NSF laws}
\end{equation}%
where $\mu $, $\mu _{b}$ and
$\kappa/ \mathrm{R}$ are positive phenomenological
coefficients, identified as the viscosity, the bulk viscosity and the thermal
conductivity of the gas, respectively. 
\subsection{Coupled Constitutive Relations from the Gibbs relation}

To obtain second-order coupled constitutive relations, the entropy flux $%
\Psi $ entails all the quadratic vector terms in $\mathbf{q}$, $\mathbf{%
\sigma }$, and $\Pi $, i.e.,%
\begin{equation}
\mathbf{\Psi }=\frac{\mathbf{q}}{\theta }-\alpha _{0}%
\frac{\mathbf{\sigma \cdot q}}{p\theta }-\beta _{0}\frac{\Pi \mathbf{q}}{%
p\theta }\text{,}  \label{Extended entropy flux}
\end{equation}%
 where $\alpha _{0}$, $\beta _{0}$ are phenomenological dimensionless coefficients. The ansatz (\ref{Extended entropy flux}) is motivated from the entropy flux of 14 moments theory \citep{pavic2013maximum}, in which $\alpha_0=\beta_0 = 2/(5+\delta)$. %; these values will be used throughout this article.
    
    % The ansatz regarding entropy flux above would be employed in the context of polyatomic gases.
    The internal degrees of freedom of such polyatomic matter have been addressed in literature as additional energy variables, see for example, \citep{pavic2013maximum}. Literature regarding such approaches appeared as early as 1967 \citep{kogan}
    and has been used effectively thereafter in macroscopic modeling of polyatomic gases \citep{liumueller,  pavicsimic, arima}, producing the correct caloric equation of state for polyatomic gases. Of course  the exact values for the phenomenological coefficients ($\alpha_0$, $\beta_0$) appearing in (\ref{Extended entropy balance equation}) depend on intricate nature of energy transfer mechanism  among internal and transnational degrees  of freedom. Nevertheless, these can also be obtained through asymptotic reduction of the kinetic equation, which we shall briefly discuss after establishing the second-law inequality.   
    % Macroscopic modeling of polyatomic gases has nonetheless been further investigated in literature. It has been partially treated by several  approaches for internal degrees of freedom other than as communicable internal energy during the collisions. However, the progress along such other lines of investigation has remained limited \citep{pavicsimic}. As a result, even recent literature addresses internal degrees of freedom of polyatomic gases in a manner similar to the one proposed by Kogan, for example \citep{arima}. 

Substitution of the extended entropy flux $\mathbf{\Psi}$ from the
last equation in (\ref{General balance equation}), and balancing the terms
give an extended second law%
\begin{equation}
\frac{\partial \left( \rho \eta \right) }{\partial t}+\bigtriangledown \cdot %
\left[ \rho \eta \mathbf{v+}\frac{\mathbf{q}}{\theta }-\frac{\alpha _{0}}{%
p\theta }\mathbf{\sigma \cdot q}-\frac{\beta _{0}}{p\theta }\Pi \mathbf{q}%
\right] =\Sigma \text{,}  \label{Extended entropy balance equation}
\end{equation}%
where the entropy production term now reads%

\begin{multline}
\Sigma =-\frac{\mathbf{\sigma }}{\theta }:\left\langle \bigtriangledown 
\mathbf{v}+\frac{\alpha _{0}}{p}\left\{\bigtriangledown \mathbf{q}-\frac{\mathbf{q} \bigtriangledown \left(p\theta \right)}{2 p \theta} \right\}\right\rangle  
%\notag \\
-\frac{\Pi }{\theta }\left[ \bigtriangledown \cdot \mathbf{v}+\frac{\beta
_{0}}{p}\left\{\bigtriangledown \cdot \mathbf{q}-\frac{\mathbf{q} \cdot\bigtriangledown \left(p\theta \right)}{2 p \theta} \right\}\right]   
\notag \\
-\frac{\mathbf{q}}{\theta ^{2}}\cdot \left[ \bigtriangledown \theta +\frac{%
\alpha _{0}}{\rho }\left\{ \bigtriangledown \cdot \mathbf{\sigma }-\frac{\mathbf{%
\sigma }\cdot \bigtriangledown \left(p\theta \right)}{2 p \theta}\right\}  +\frac{\beta _{0}}{\rho }\left\{ \bigtriangledown \Pi -\frac{\Pi 
\bigtriangledown \left(p \theta\right)}{2 p \theta}\right\} \right] \text{.}
\label{CCR entropy production}
\end{multline}%
Again, the entropy production forms a bilinear form as a sum of products of
the thermodynamic fluxes and generalized thermodynamic forces. 
% In (\ref%
% {CCR entropy production}) four dimensionless (arbitrary) coefficients ($%
% \alpha _{1,2}$, $\beta _{1,2}$, $\alpha _{1,2}^{\ast }:=1-\alpha _{1,2}$, $%
% \beta _{1,2}^{\ast }:=\beta _{1,2}$) are introduced, which appear as weights
% for the different force-flux pairs contributions to the entropy production;
% their values can be obtained from comparison to results from kinetic theory.
Once again, in order to guarantee the positivity of the entropy production, a
linear flux-force relationship is assumed, which leads to the following
coupled constitutive relations%
\begin{eqnarray}
\mathbf{\sigma } =-2\mu \left\langle \underline{\bigtriangledown 
\mathbf{v}}+\frac{\alpha _{0}}{p}\left\{\bigtriangledown \mathbf{q}-\frac{\mathbf{q} \bigtriangledown \left(p\theta \right)}{2 p \theta} \right\}\right\rangle \text{, }\Pi  =-\mu _{b}\left[ \underline{\bigtriangledown \cdot \mathbf{v}}+\frac{\beta
_{0}}{p}\left\{\bigtriangledown \cdot \mathbf{q}-\frac{\mathbf{q} \cdot\bigtriangledown \left(p\theta \right)}{2 p \theta} \right\}\right] \label{CCR stress balance}
% \\
% \Pi  &=&-\mu _{b}\left[ \underline{\bigtriangledown \cdot \mathbf{v}}+\frac{\beta
% _{0}}{p}\left\{\bigtriangledown \cdot \mathbf{q}-\frac{\mathbf{q} \cdot\bigtriangledown \left(p\theta \right)}{2 p \theta} \right\}\right]\text{,}
% \label{CCR dynamic pressure}
\end{eqnarray}%
and
\begin{eqnarray}
\mathbf{q} =-\kappa \left[ \underline{\bigtriangledown \theta }+\frac{%
\alpha _{0}}{\rho }\left\{ \bigtriangledown \cdot \mathbf{\sigma } -\frac{\mathbf{\sigma }\cdot \bigtriangledown
\left(p\theta \right)}{2p\theta}\right\}  +\frac{\beta _{0}}{\rho }\left\{ \bigtriangledown \Pi -\Pi \frac{\bigtriangledown \left(p \theta \right)}{2p\theta}\right\} \right] \text{.}  \label{CCR heat balance}
\end{eqnarray}%
The stress (\ref{CCR stress balance}) given by the CCR theory entails the hydrodynamic
stress (underlined term) and the non-Newtonian second order corrections.
Similarly, the heat flux equation (\ref{CCR heat balance}) contains
Fourier's law (underlined term), along with the second-order corrections to
the heat flux. 
% The CCR theory also brings forth a coupling between the heat
% flux and the dynamic pressure via the coefficient $\beta _{0}$. Clearly, the CCR (\ref{CCR stress balance}--\ref{CCR heat balance}) reduce to the
% classical NSF closure (\ref{NSF laws}) for $\alpha _{0}=\beta _{0}=0$; the non-zero
% values for these coefficients introduce second-order contributions to the
% NSF closure. 

 In order to study non-equilibrium effects in rarefied polyatomic gases,  \citet{RahimiStruchtrup2014} introduced a set of 36 moment equations and further utilized the order of magnitude method to
identify the leading order terms of all the moments appearing in the 36
moment theory. In their work the balance equations for the dynamic
temperature $\vartheta =\Pi /\rho $ reads (eq. 35d in \citep{RahimiStruchtrup2014})%
\begin{equation}
\underline{\rho \frac{D\vartheta }{Dt}+\frac{\delta }{3}\frac{2}{3+\delta }(
\Pi \nabla \cdot \mathbf{v}+\sigma
:\nabla \mathbf{v})+\frac{\delta }{3}\frac{2}{5+\delta }\nabla \cdot \Gamma }+%
\frac{\delta }{3}\frac{2}{3+\delta }\left( p\nabla \cdot \mathbf{v+}\frac{2}{%
5+\delta }\nabla \cdot \mathbf{q}\right)  =-\frac{1}{\tau
_{int}}\Pi 
\label{Balance equation for vartheta}
\end{equation}
where $D/Dt\ $denotes the material time derivative, $\tau _{int}$ is the
relaxation time for the dynamic temperature, and one defines $\Gamma =%
\mathbf{q}^{tr}-\frac{5}{\delta }\mathbf{q}^{in}$; $\mathbf{q}^{tr}$ and $%
\mathbf{q}^{in}$ being the translational and the internal heat fluxes,
respectively. Furthermore, using the Chapman-Enskog expansion (in $\tau _{in}
$), is was shown in the same article, that the underlined terms in last equation are
higher order compared to the others, which can be assumed small in first order approximations. Neglecting these higher order terms, one gets the leading order constitutive relation for $\Pi $, as 
\begin{equation}
\Pi ^{(1)}=-\tau _{int}\frac{\delta }{3}\frac{2}{3+\delta }\left( p\nabla
\cdot \mathbf{v}+\frac{2}{5+\delta }\nabla \cdot \mathbf{q}\right) \text{.}
\end{equation}
Comparing last equation with (\ref{CCR stress balance}$_2$), one immediately recognizes 
\[
\tau _{int}=\frac{3}{\delta }\frac{3+\delta }{2}\frac{\mu _{b}}{p}\text{ and 
}\beta _{0}=\frac{2}{5+\delta }\text{.}
\]%
Similarly, the Chapman-Enskog expansion---via introducing a smallness
parameter $\varepsilon $, and assuming $\epsilon \sim O\left( \varepsilon
^{1/2}\right) $, $\tau _{tr}\sim O\left( \varepsilon ^{\alpha /2}\right) $, $%
\tau _{in}\sim O\left( \varepsilon ^{1/2}\right) $ and $0<\alpha <1$---of
the balance equation for heat flux follows (from eq. 35f in \citep{RahimiStruchtrup2014}), one again
finds $\beta _{0}=\frac{2}{5+\delta }$. Similar, arguments based upon the Chapman-Enskog can be made to justify $\alpha _{0}=\frac{2%
}{5+\delta }$. Nevertheless, here we shall point out that the value for these
phenomenological coefficients may vary with different collision models;
consideration of which are beyond the scope of this article and throughout
this article we shall assume 
\begin{equation}
\alpha _{0}=\frac{2}{5+\delta }\text{,\quad }\beta _{0}=\frac{2}{5+\delta }\text{.%
}
\end{equation}
Henceforth, we shall refer to (\ref{CCR stress balance}--\ref{CCR heat balance}) as polyatomic-CCR model.

\section{Linearized and dimensionless equations}
In this section, the equations, which we consider are dimensionless and
linearized by introducing small perturbations from their values in a
reference rest state defined by a constant pressure $p_{0}$ and
a constant temperature $\theta _{0}$. The relations between the field
variables and their dimensionless deviations (denoted with hat symbols) from
their reference rest state are given as%
\begin{eqnarray*}
p =p_{0}\left( 1+\hat{p}\right) \text{, }\theta =\theta _{0}\left( 1+\hat{%
\theta}\right) \text{, }\mathbf{v=}\sqrt{\theta _{0}}\mathbf{\hat{v}}\text{, } 
\mathbf{\sigma } =p_{0}\mathbf{\hat{\sigma}}\text{, }
\Pi = p_{0}\hat{\Pi}%
\text{, }\mathbf{q}=p_{0}\sqrt{\theta _{0}}\mathbf{\hat{v}}\text{, and }\mathbf{x%
}=L\mathbf{\hat{x}}\text{,}
\end{eqnarray*}%
where $L$ is a characteristic length scale. The linearized and stationary (%
$\partial /\partial t\equiv 0$) conservation laws are given by

\begin{equation}
\bigtriangledown \cdot \mathbf{\hat{v}} =0\text{,}\quad
\bigtriangledown \cdot \left[ \left( \hat{p}+\hat{\Pi}\right) \mathbf{I+\hat{%
\sigma}}\right] =0\quad\text{and}\quad 
\bigtriangledown \cdot \mathbf{\hat{q}} =0\text{,}
\label{Linear conservation laws}
\end{equation}%
and the linearized stress, dynamic pressure and heat flux are specified as 

\begin{equation}
\mathbf{\hat{\sigma}} =-2\mathrm{Kn}\left\langle \bigtriangledown \mathbf{%
\hat{v}}+\frac{2}{\delta +5}\bigtriangledown \mathbf{\hat{q}}\right\rangle 
\text{,}\quad\hat{\Pi}=0\quad\text{and}\quad
\mathbf{q} =-\frac{\mathrm{Kn}c_{p}}{\mathrm{Pr}}\left[ \bigtriangledown \hat{
\theta}+\frac{2}{\delta +5}\bigtriangledown \cdot \mathbf{\hat{\sigma}}
\right] \text{.}  \label{Linear constitutive}
\end{equation}
In (\ref{Linear constitutive}), the Knudsen number appears as the scaled viscosity $\mathrm{Kn}=\mu _{0}\sqrt{\theta _{0}}/\left(p_{0}L\right)$,
and $\mathrm{Pr}= c_{p}\mu_{0}/\kappa$ is the Prandtl number. Here, it is assumed that all the field
variables are measured relative to their equilibrium values at infinity,
hence all perturbations must vanish as $|\mathbf{\hat{x}}|\rightarrow \infty $.

%%%%%%%%%%%%%%%%%%%%%%%%%%%%%%%
%%%%%%%%%%%%%%%%%%%%%%%%%%%%
%%%%%%%%%%%%%%%%%%%%%%%%%%%%
% I suggest we cut down the following subsection as I have done below%

\subsection{Mathematical similarities between the CCR and NSF equations}

It is worth noting that the steady, linearized CCR equations (\ref{Linear
conservation laws}
--\ref{Linear constitutive}) can be converted to the
Stokes-Fourier like system by introducing the following auxiliary variables, 
% $\mathbf{\hat{V}} :=\mathbf{\hat{v}}+\alpha_0\bigtriangledown\mathbf{\hat{q}}$ and $\hat{\Theta} :=\hat{\theta}-\alpha_0\hat{p}$.
 \begin{equation}
 \mathbf{\hat{V}} :=\mathbf{\hat{v}}+\frac{2}{\delta +5}\bigtriangledown\mathbf{\hat{q}}\text{\quad and \quad}
\hat{\Theta} :=\hat{\theta}-\frac{2}{\delta +5}\hat{p}\text{.}
\label{Aux variables}
 \end{equation}
As a result, the linearized conservation laws (\ref{Linear conservation laws}) and (\ref{Linear constitutive}) turn into
\begin{equation}
\bigtriangledown \cdot \mathbf{\hat{V}}=0\text{, }\bigtriangledown \hat{p}=
\mathrm{Kn}\Delta \mathbf{\hat{V}}\quad\text{and}\quad\Delta \hat{\Theta}=0\text{,}
\end{equation}%
% {\color{blue} Removed the troublesome para which was below. Put two percent signs in the LaTeX file instead of one whereever I commented out phrases.}
which are mathematically the same as the Stokes equations in $\{\mathbf{\hat{V}}, \hat{p}\}$
and the Laplace equation in $\hat{\Theta}$; however, now the Stokes equations
are coupled with the Laplace equation via auxiliary variables $\mathbf{\hat{V}}$ and $\hat{\Theta} $. Nevertheless, from a mathematical point of view, the classical
analytic solutions and techniques from potential theory \citep{batchelor2000introduction} for the Stokes and
Laplace equations can be carried over to the slow, steady-state polyatomic-CCR model, as we show in next section.
%In order to assess the accuracy of newly derived CCR equations, we shall
% apply these equations to some classical boundary value problems and compare
% the results with experiments/other theoretical models from literature.
%%%%%%%%%%%%%%%%%%%%%%%%%%%%%%%

% It is worth noting that the steady, linearized CCR equations (\ref{Linear
% conservation laws}--\ref{Linear constitutive}) can be converted to the
% Stokes-Fourier like system by introducing auxiliary variables ($\mathbf{\hat{V}} :=\mathbf{\hat{v}}+\alpha_0 \mathbf{\hat{q}}$ and $\hat{\Theta} :=\hat{\theta}-\alpha_0\hat{p}$). Potential theory can thus provide classic analytical solutions which
% can be carried over to slow steady-state polyatomic-CCR model as we show in the next section.

% {\color{blue} I think we should put in explicitly how the Stokes Fourier like system is obtained. Could you please detail it a bit here?} 

In order to assess the accuracy of newly derived CCR model, we shall apply these equations to two classical boundary value problems (i) slow rarefied gas flow over a spherical particle, for which a closed form solution can be obtained, and (ii)  slow rarefied gas flow over a doublet, which is solved numerically using the method of fundamental solution (MFS) \citep{lockerby_collyer_2016, RSCLS2021}.

\section{Slow rarefied gas flow over a spherical particle}
\label{sec: Slow rarefied gas flow}

First, we consider the case of a slow rarefied gas flow around a
spherical particle of radius $L$. For this problem, solutions of the NSF 
equations are  available \citep{batchelor2000introduction},
giving the pressure $\hat{p}$, the velocity in radial direction $\hat{u}_{r}$ and the 
azimuthal directions $\hat{u}_{\phi }$ as

\begin{equation}
\hat{p}=\mathrm{Kn}\frac{a_{1}}{\hat{r}^{2}}\cos \phi\text{,  } \hat{u}_{r} =\hat{U}_{\infty }\left[ 1+\frac{a_{1}}{\hat{r}}+\frac{a_{2}}{%
\hat{r}^{3}}\right] \cos \phi\text{ and }
\hat{u}_{\varphi } =-\hat{U}_{\infty }\left[ 1+\frac{a_{1}}{2\hat{r}}-%
\frac{a_{2}}{2\hat{r}^{3}}\right] \sin \phi\text{,}
\label{velocity problem}
\end{equation}%
in the spherical coordinate system. Here, $\hat{U}%
_{\infty }$($=U_{\infty }/\sqrt{\theta _{0}}$) is the far-field velocity of
the gas and $\hat{r}$ ({\color{blue}$=r/L$}) is the dimensionless radial distance, $\hat{r}$. The temperature and the heat flux are obtained as 
\begin{equation}
\hat{\theta}=%
\frac{c_{1}}{\hat{r}^{2}}\cos \phi \text{, } \hat{q}_{r} = \frac{2 c_p \mathrm{Kn}}{\mathrm{Pr}}\frac{c_1-\alpha_0 a_1 \mathrm{Kn}}{\hat{r}^3}\cos \phi\text{ and  }\hat{q}_{\phi} = \frac{c_p \mathrm{Kn}}{\mathrm{Pr}}\frac{c_1-\alpha_0 a_1 \mathrm{Kn}}{\hat{r}^3}\sin \phi
\label{temperature problem}
\end{equation}%
The integration constants ($a_{1}$, $a_{2}$ and $c_{1}$) are obtained by
applying appropriate boundary conditions. The Stokes formula for the drag force
exerted on a spherical particle due to the gas flow reads $\hat{F}_{S}=-6\pi \mathrm{Kn}\hat{U}_{\infty }$, 
% \begin{equation}
% \hat{F}_{S}=-6\pi \mathrm{Kn}\hat{U}_{\infty }\text{,}  \label{Stokes Drag}
% \end{equation}%
which is obtained by applying the classical no-slip and no-jump boundary
condition at $\hat{r}=1$, i.e., $\hat{u}_{r}=\hat{u}_{\phi }=0$ and $\hat{%
\theta}=0$ to the NSF solutions, i.e., equations (\ref{velocity problem}--\ref{temperature problem}) with $\alpha_0=0$. The Stokes' drag formula is valid for $\mathrm{Kn}\rightarrow
0 $. Likewise, the NSF solutions with the standard first-order slip and
temperature jump boundary conditions, i.e., 
\begin{equation}
\hat{u}_{r}=0\text{, }\hat{u}%
_{\phi }=-\sqrt{\frac{\pi}{2} }\hat{\sigma}_{r\phi }\quad\text{and}\quad\hat{\theta}=-\frac{2}{(4+\delta)}\sqrt{\frac{\pi}{2} }\hat{q}_{r}\text{,}
\label{first-order BCs}
\end{equation}%
yield drag force $\hat{F}_{D} = \hat{F}_{S} \left(2\mathrm{Kn}+\sqrt{2/\pi}\right)/\left(3
\mathrm{Kn}+\sqrt{2/\pi}\right)$. In (\ref{first-order BCs}), the accommodation coefficients are taken to be unity \citep{SharipovData}.
%tangential and viscous slip and the thermal slip coefficients are taken to be unity 
% \begin{equation}
% \frac{\hat{F}_{D}}{\hat{F}_{S}}=\frac{2\mathrm{Kn}+\eta _{\mathrm{VS}}}{3%
% \mathrm{Kn}+\eta _{\mathrm{VS}}}\text{.}  \label{NSF drag}
% \end{equation}%

The second order boundary conditions which shall be used for the polyatomic-CCR solutions (\ref{velocity problem}--\ref{temperature problem}), read 
\begin{equation}
\hat{u}_{r}=0\text{, }\hat{u}%
_{\phi }=-\sqrt{\frac{\pi}{2} }\hat{\sigma}_{r\phi }-\underline{\frac{3}{4}}\hat{q}%
_{\phi }\quad\text{and }\quad\hat{\theta}=-\underline{\frac{2}{(4+\delta)}\sqrt{\frac{\pi}{2}} }\hat{q}_{r}-\frac{3}{4}\hat{\sigma}%
\label{2nd boundary condition}
_{rr }\text{.}
\end{equation}% 
Here, the first boundary condition (\ref{2nd boundary condition}$_1$) represents the non-penetration condition on a solid wall. The slip boundary condition (\ref{2nd boundary condition}$_2$) defines the velocity slip at the wall, which relates the tangential gas velocity slip  the tangential shear stress and heat flux. The slip boundary condition, when compared to the second-order boundary conditions from the literature \citep{DEISSLER1964681, Lockerby2004,reviewslip2012} gives the first-order and second-order slip coefficients as $A_1=1$ and $A_2 =\frac{3}{2\pi }\frac{1}{%
\Pr}) =0.682$; see Appendix \ref{sec: Asymptotic analysis of the boundary conditions} for details. The third boundary condition (\ref{2nd boundary condition}$_3$) is the  temperature-jump boundary condition, where the temperature jump coefficient  $\zeta _{T}=\frac{\sqrt{\pi }}{2}\frac{2}{4+\delta }\frac{c_{p}}{\Pr }$ \citep{SharipovData} and the accommodation coefficient being
one.   An additional term in (\ref{2nd boundary condition}$_3$) appears as a consequence of the Onsager symmetry \citep{RGS2018,rana_gupta_sprittles_torrilhon_2021}.

 It should be noted here that values for the slip and jump coefficients vary in literature \citep{SharipovData} and one may perform asymptotic expansions---taking into account the so-called Knudsen layer corrections \citep{rana_struchtrup_2016}---to fit theory/experimental results. Indeed, in (\ref{2nd boundary condition}$_2$) the coefficient $3/4$ (which is equivalent of taking an appropriate value for the second-order slip coefficient $A_2$) was chosen so that it gives good fit for the drag
coefficient for a sphere; see Appendix \ref{sec: Asymptotic analysis of the boundary conditions} for further details along with the sensitivity analysis of the results on slip/jump coefficients.  In particular, the considered choice of $A_2$ significantly overestimate  the value for   thermal slip coefficient \citep{SharipovData}. Nevertheless, for the problems considered in this paper, the effects of thermal slip coefficient are negligible. 

Substitution of the CCR solutions (\ref{velocity problem}--\ref{temperature problem}) in (\ref{2nd boundary condition}) gives integration constants ($a_{1}$, $a_{2}$ and $c_{1}$), which in turn provide a closed form expression for the drag as
\begin{equation}
    \hat{F}_{D} = \hat{F}_{S} \left(1+kn\frac{1+\left(\frac{3kn}{2\pi }\frac{1+4\frac{3\delta +19}{\delta +5}kn}{\Pr +2%
\frac{\delta +5}{\delta +4}kn}\right)}{1+kn\left( 2+\frac{9kn}{4\pi }\frac{3-\delta }{\Pr +%
2\frac{\delta +5}{\delta +4}kn}\right) }\right)^{-1}
\end{equation} 
% \begin{equation}
%     \hat{F}_{D} = \hat{F}_{S} \left(1+kn\frac{1+\frac{3}{2\pi }\sqrt{\frac{\pi }{2}%
% }Kn\frac{1+4Kn\sqrt{\frac{\pi }{2}}\frac{3\delta +19}{\delta +5}}{\Pr +2%
% \sqrt{\frac{\pi }{2}}Kn\frac{\delta +5}{\delta +4}}}{1+\sqrt{\frac{\pi }{2}}%
% Kn\left( 2+\frac{9}{4\pi }\sqrt{\frac{\pi }{2}}Kn\frac{(3-\delta )}{\Pr +%
% \sqrt{\frac{\pi }{2}}Kn\frac{2(\delta +5)}{\delta +4}}\right) }\right)^{-1}
% \end{equation}
where, for brevity, we have introduced a modified Knudsen number $kn = \sqrt{\frac{\pi }{2}}\mathrm{Kn}$. Clearly, $\hat{F}_{D} \rightarrow \hat{F}_{S}$ as  $\mathrm{Kn}\rightarrow 0$ and $\hat{F}_{D}/\hat{F}_{S}\rightarrow 0$ as $\mathrm{Kn}\rightarrow \infty$, hence the continuum and the free-molecular regimes are recovered, qualitatively.

In Fig.~(\ref{fig:dragsphere}) the drag force $\hat{F}_D$ normalized with Stokes' drag $\hat{F}_S$ is shown as
a function of the Knudsen number. The results obtained from the polyatomic-CCR model (solid blue),
first-order NSF (dashed red) and Stokes' formula (thin green)
 are compared with the experimental results (black dot-dashed lines and symbols) given by \citet{ALLEN1982537}. 
 
 For larger values of the
Knudsen number, the force decreases. Notably, for the NSF with the first-order
boundary conditions the drag force reaches a finite value ($2/3$) as $\mathrm{Kn}%
\rightarrow \infty $, which contradicts experimental observations.  The polyatomic-CCR results
match with the experiment quite remarkably even for large value of the Knudsen number. In the above computations,
transport properties are taken to be for dry air, i.e., $\Pr =0.7$ and  $\delta=2$, which gives  the ratio of the specific heats $\gamma = 1.4$---a value consistent with air/nitrogen at moderate temperatures.

\section{Slow rarefied gas flow over a doublet}
\label{sec: Slow rarefied gas flow doublet}
 In this section we shall consider the non-trivial case of flow over a doublet for which we shall apply the method of fundamental solution (MFS) technique. However, before that, we first formulate the  Green functions associated with the polyatomic-CCR model (\ref{Linear conservation laws}--\ref{Linear constitutive}). 
\subsection{Green's functions and the method of fundamental solutions}
Green's functions for the Stokes equations and the Laplace equations are
obtained as a response of the fluid with regard to a point body force $\mathbf{%
\hat{f}}$ in the momentum equation (Stokeslet) and a
heat source of strength $\hat{g}$ (Thermlet) in the energy
balance equation \citep{lockerby_collyer_2016, RSCLS2021}. Green's functions associated with the polyatomic-CCR equations
follow from the Stokeslet and Thermlet, as 
\begin{subequations}
\label{Gradlet solutions}
\begin{eqnarray}
\mathbf{\hat{v}}{\left( \mathbf{\hat{r}}\right) } &=&\frac{1}{8\pi }\left(\frac{\mathbf{\hat{r}\hat{r}}%
}{|\mathbf{\hat{r}}|^{3}}+\frac{\mathbf{I}}{|\mathbf{\hat{r}}|}\right)\cdot \mathbf{\hat{f}}+\underline{\frac{%
3\alpha _{0}^{2}c_{p}\mathrm{Kn}^{2}}{4\pi \Pr |\mathbf{\hat{r}}|^{5}}\langle\mathbf{\hat{r}\hat{r}}\rangle\cdot \mathbf{\hat{f}}}\text{, } 
  \label{Gradlet pressure} \\
\hat{p}{\left( \mathbf{\hat{r}}\right) } &=&\frac{\mathrm{Kn}}{4\pi }\frac{\mathbf{%
\hat{f}\cdot \hat{r}}}{|\mathbf{r}|^{3}}\quad\text{and }\quad\mathbf{\hat{\sigma} }{\left( \mathbf{\hat{r}}\right) } =\frac{3\mathrm{Kn}}{%
4\pi |\mathbf{\hat{r}}|^{5}}\left( \mathbf{\hat{f}\cdot \hat{r}}+\underline{2\alpha _{0}\mathrm{Kn}%
\hat{g}}\right) \langle\mathbf{\hat{r}\hat{r}}\rangle\text{,}
\label{Gradlet stress}
\end{eqnarray}%
\end{subequations}
where $\mathbf{\hat{r}}=%
 \mathbf{\hat{x}-\hat{x}}^{s}$ is the position of an arbitrary point $
 \mathbf{\hat{x}}$ from the computational domain relative to the
 singularity point located at $\mathbf{\hat{x}}^{s}$ (a point outside of the
 computational domain).
%where $\alpha _{0}=1/c_{p}$.

% The Oseen-Burger tensor, $\mathbf{J}{\left( 
% \mathbf{\hat{r}}\right) }$ and the third decaying harmonic tensor, $\mathbf{K%
% }{\left( \mathbf{\hat{r}}\right) }$ are defined as 
% \end{subequations}
% \begin{equation}
% \mathbf{J}{\left( \mathbf{\hat{r}}\right) } =\frac{\mathbf{\hat{r}\hat{r}}%
% }{|\mathbf{\hat{r}}|^{3}}+\frac{\mathbf{I}}{|\mathbf{\hat{r}}|}\text{, and }
% \mathbf{K}{\left( \mathbf{r}\right) } =\frac{\mathbf{\hat{r}\hat{r}}}{|%
% \mathbf{\hat{r}}|^{5}}-\frac{1}{3}\frac{\mathbf{I}}{|\mathbf{\hat{r}}|^{3}}\text{,}
% \label{third decaying harmnonic}
% \end{equation}%
% where $\mathbf{\hat{r}}=%
% \mathbf{\hat{x}-\hat{x}}^{s}$ is the position of any arbitrary point $%
% \mathbf{\hat{x}}$ from the computational domain relative to the
% singularity point located at $\mathbf{\hat{x}}^{s}$ (a point outside of the
% computation domain).

The temperature $\hat{T}$ and heat flux $\mathbf{\hat{q}}$ are obtained as 
\begin{equation}
\hat{T}{\left( \mathbf{\hat{r}}\right) } =\frac{\Pr }{4\pi c_{p}}\frac{\hat{g}}{|%
\mathbf{\hat{r}}|}\quad\text{and } \quad 
\mathbf{\hat{q}}{\left( \mathbf{\hat{r}}\right) } =\frac{\mathrm{Kn}}{4\pi }%
\frac{\hat{g}}{|\mathbf{\hat{r}}|^{3}}\mathbf{\hat{r}}-\underline{\frac{%
3\alpha _{0}c_{p}\mathrm{Kn}^{2}}{4\pi \Pr |\mathbf{\hat{r}}|^{5}}\langle\mathbf{\hat{r}\hat{r}}\rangle\cdot \mathbf{\hat{f}}}\text{.}  \label{Gradlet  solutions2}
\end{equation}
In the polyatomic-CCR Green's functions (\ref{Gradlet solutions}--\ref{Gradlet solutions2}%
), the coupling between the Stokeslet and Thermlet occurs due to the underlined
term. Obviously, for the classic NSF equations ($\alpha _{0}=0$), such
coupling is absent and one obtains classical Stokeslet and Thermlet.  

\subsection{Drag on a doublet}
\begin{figure}
\centering
\begin{subfigure}{.55\textwidth}
  \centering
  \includegraphics[width=\linewidth]{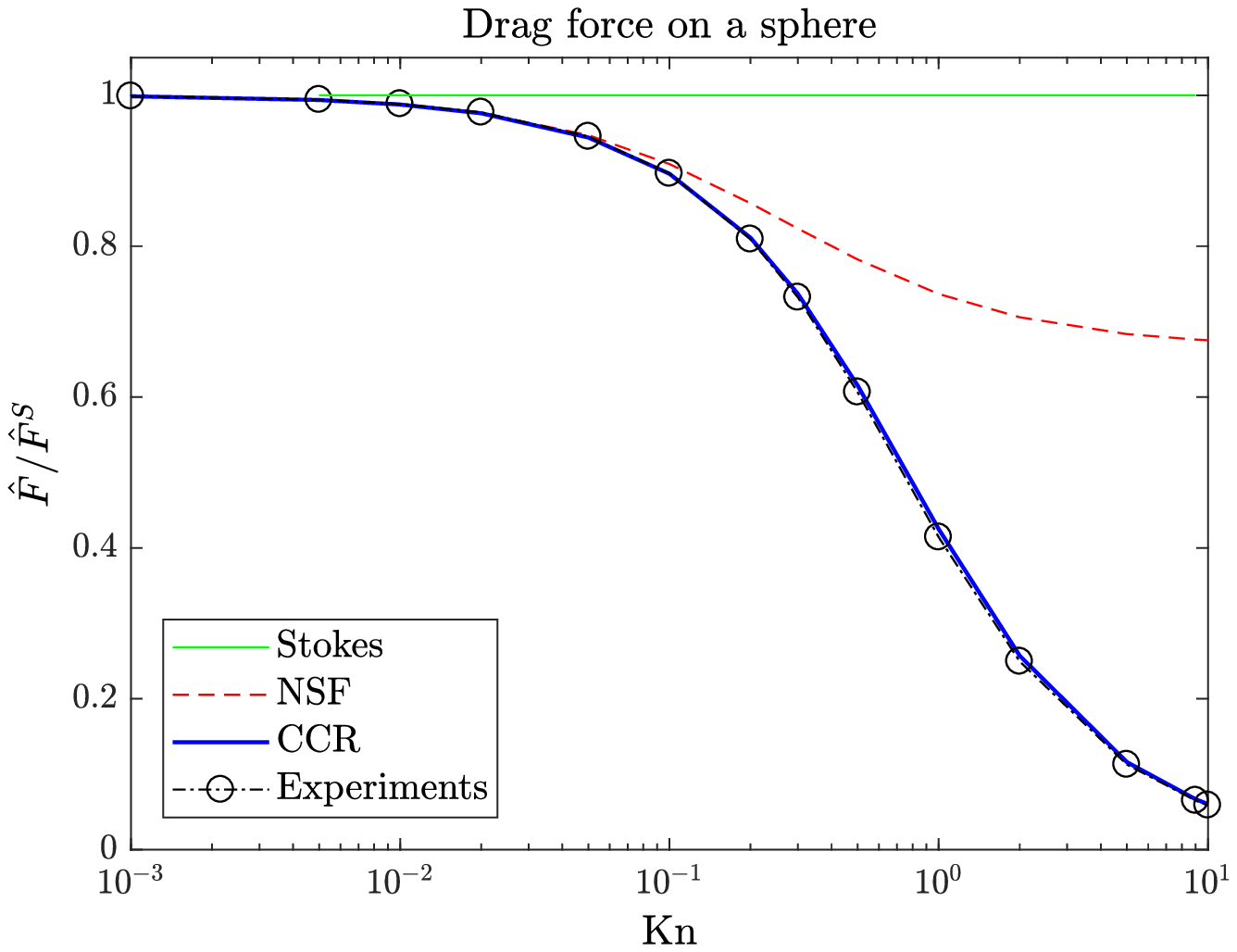}
  \caption{The normalised drag force  (with Stokes' drag) on a sphere  against Knudsen number. The results computed from the NSF with first-order slip and the polyatomic-CCR model are compared with the experimental data by \citet{ALLEN1982537}.}
  \label{fig:dragsphere}
\end{subfigure}%
\begin{subfigure}{.55\textwidth}
  \centering
  \includegraphics[width=\linewidth]{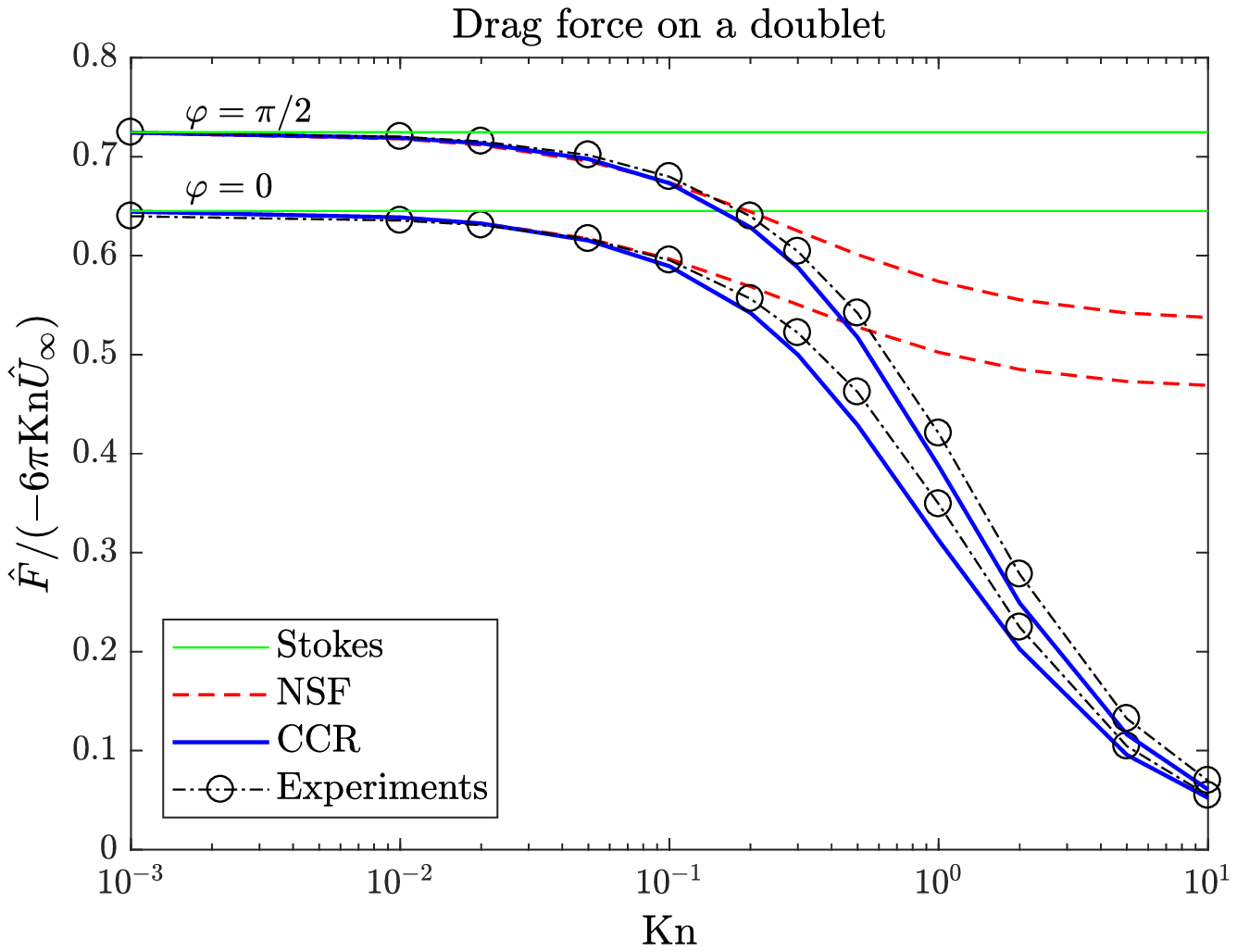}
  \caption{The drag force on a doublet against Knudsen number.  Two different flow  conditions are assumed, i.e, flow parallel to its line of centres ($\varphi = 0$) and
 flow perpendicular to its line of centres ($\varphi = \pi/2$). The experimental data are taken from \citet{Cheng1988}.}
  \label{fig:dragdoublet}
\end{subfigure}
\caption{}
\label{fig:test}
\end{figure}

In Fig.~(\ref{fig:dragdoublet}), we show the drag force on a doublet
using the MFS technique developed by \citet{lockerby_collyer_2016, RSCLS2021} for the case of monoatomic gases. The essential idea behind MFS is to decompose the solutions of the linearized partial differential equations as superposition of the Green's functions, where the weights are determined such that the boundary conditions are satisfied at some boundary nodes (collocation points).

In particular, we study a doublet moving with an angle of $%
\varphi $ between the flow direction and the line joining the centers of the
doublet. This is an interesting case to be considered because (i) it is a
full 3D problem and no closed form analytic solution exits (ii) experimental
results are available for this problem in the literature \citep{Cheng1988} for $\varphi =0$
and $\varphi =\pi /2$ \citep{Cheng1988}. The overall
agreement of the polyatomic-CCR model's drag force with the experimental data is excellent in both the cases ($\varphi = 0,\pi/2$)  and as before, the NSF theory over-predicts the drag. 
% Thus, the polyatomic-CCR model along with the MFS technique can provide provide a fully
% predictive and accurate simulation-framework for aerodynamics of complex shaped UFPs. 

% and compare the results with those obtained from the classical NSF theory and the experimental results reported in [ref].
%   A comparison of the steady-state linearized CCR-polyatomic equations (\ref{Linear conservation laws}--\ref{Linear constitutive}) with their monoatomic counterparts \citep{RSCLS2021, RGS2018} reveals that the only difference between two models appear through $\mathrm{Pr}$, $c_p$, $c_v$, $\alpha_0$ and $\beta_0$, where last two values being $2/5$ and $0$ for monoatomic gases, respectively. It is due to the fact that dynamic pressure is identically zero (from eq.~\ref{Linear constitutive}$_2$). For nonlinear or transient process, the bulk pressure may no be zero  
  
So far, we have considered steady-state slow rarefied flow conditions for which the polyatomic-CCR model can be transformed into the NSF-like equations and analytic/semi-analytic approaches can be utilized. For the considered cases, the dynamic temperature identically vanishes and the polyatomic nature of the gas only appear through $\mathrm{Pr}$, $\mu$, $c_p$, $c_v$, $\alpha_0$ and $\beta_0$, where last two values being $2/5$ and $0$ for monoatomic gases, respectively. For non-linear (for example, shocks in hyper-sonic flows) or transient process (for example, wave propagation and Rayleigh-Brillouin scattering in polyatomic gases), other polyatomic effects might manifest itself.
% Henceforth, in order to assess the effects of non-linear terms, in next section, we shall use the polyatomic-CCR model to compute the shock structure of nitrogen and compare the results with those obtained from the classical NSF theory and the experimental results reported in [ref].
\section{Normal shock structure in nitrogen}
\label{sec: normal shock structure in nitrogen}
Finally, in order to assess the effects of non-linear terms,  we shall now use the polyatomic-CCR model to compute the shock structure in nitrogen gas. The normal shock structure problem is modelled as a one-dimensional problem so that the field variables (velocity, stress tensor and heat flux vector, etc.) have only non-zero $x$-component.  In the frame of
reference of the moving shock (in the $x$-direction), the gas upstream and downstream is assumed to be in equilibrium state with ($\rho$, $v_x$, $\theta$) = ($\rho_0$, $U_0$, $\theta_0$) and ($\rho$, $v_x$, $\theta$) = ($\rho_\infty$, $v_\infty$, $\theta_\infty$), respectively. The upstream and downstream conditions are linked to each other by the ankine-Hugoniot relations \citep{anderson2003modern}:
\begin{equation}
   \frac{\rho_{\infty}}{\rho_0} =  \frac{(\gamma +1) \text{Ma}^2}{2+\text{Ma}^2\left(\gamma  -1\right)}\text{, }\frac{\rho_{\infty}v_{\infty}}{\rho_0\sqrt{\theta_0}} = \sqrt{\gamma}\text{Ma, and }\frac{\rho_{\infty}\theta_{\infty}}{\rho_0\theta_0} = \frac{1-\gamma +2 \gamma  \text{Ma}^2}{1+\gamma}
   \label{Rankine-Hugoniot relations}
\end{equation}
where $\text{Ma} = U_0/\sqrt{\gamma \theta_0}$ is the Mach number defined in terms of the upstream condition.

The mean free
path based upon the upstream conditions can be defined as $\lambda_0 = (8 \sqrt{2}\mu_0)/\left(5\rho_0 \sqrt{\pi \theta_0}\right)$, where $\mu_0$ is the viscosity in the upstream state.
For the shock problem, the polyatomic-CCR model reduces to 
\begin{eqnarray}
\tau _{xx} &:=& \Pi+\sigma_{xx} = -\frac{4\mu }{3}\left( 1+\frac{\delta}{2\left (3+\delta\right)}Z%
\right) \left( \frac{dv_{x}}{dx}+\frac{\alpha _{0}}{p}\left\{\frac{dq_{x}}{dx}-\frac{q_x}{2p\theta}\frac{d\left(p \theta\right)}{dx}\right\}%
\right)\text{, }  \\
q_{x} &=&-\frac{\mu c_{p}}{\Pr }\left[ \frac{d\theta }{dx}+\frac{\alpha _{0}%
}{\rho }\left\{ \frac{d\left(\Pi+\sigma_{xx}\right)}{dx}-\frac{\Pi+\sigma_{xx}}{2 p\theta}\frac{d \left(p\theta\right) }{dx}%
\right\} \right].
\end{eqnarray}%
Here, for convenience, we have defined $\tau _{xx} = \Pi+\sigma_{xx}$ and introduced $Z$ as the dimensionless rotational collision number. The viscosity is obtained from the Sutherland's law and $Z$ from Parker's formula \citep{Parker1959} as
\begin{equation}
   \mu =\mu _0 \left(\frac{T}{T_0}\right)^{3/2}\frac{T_0+T_S}{T+T_S}\text{, and } Z:=\frac{3(3+\delta)\mu_b}{2\delta\mu } = \frac{Z_\infty}{1+\frac{\pi ^{3/2}}{2}  \sqrt{\frac{T^*}{T}}+\left(\frac{\pi ^2}{4}+\pi \right)\frac{T^*}{T}}
\end{equation}
where $T_s = 107 \mathrm{K}$ is the Sutherland's constant, $Z_\infty = 18.2$ and $T^{*} = 91.5 \mathrm{K}$.
The reduced system of ordinary differential equations are numerically solved using a second order central finite difference scheme with boundary conditions (\ref{Rankine-Hugoniot relations}). 

\begin{figure}
\centering
\begin{subfigure}{.33\textwidth}
  \centering
  \includegraphics[width=\linewidth]{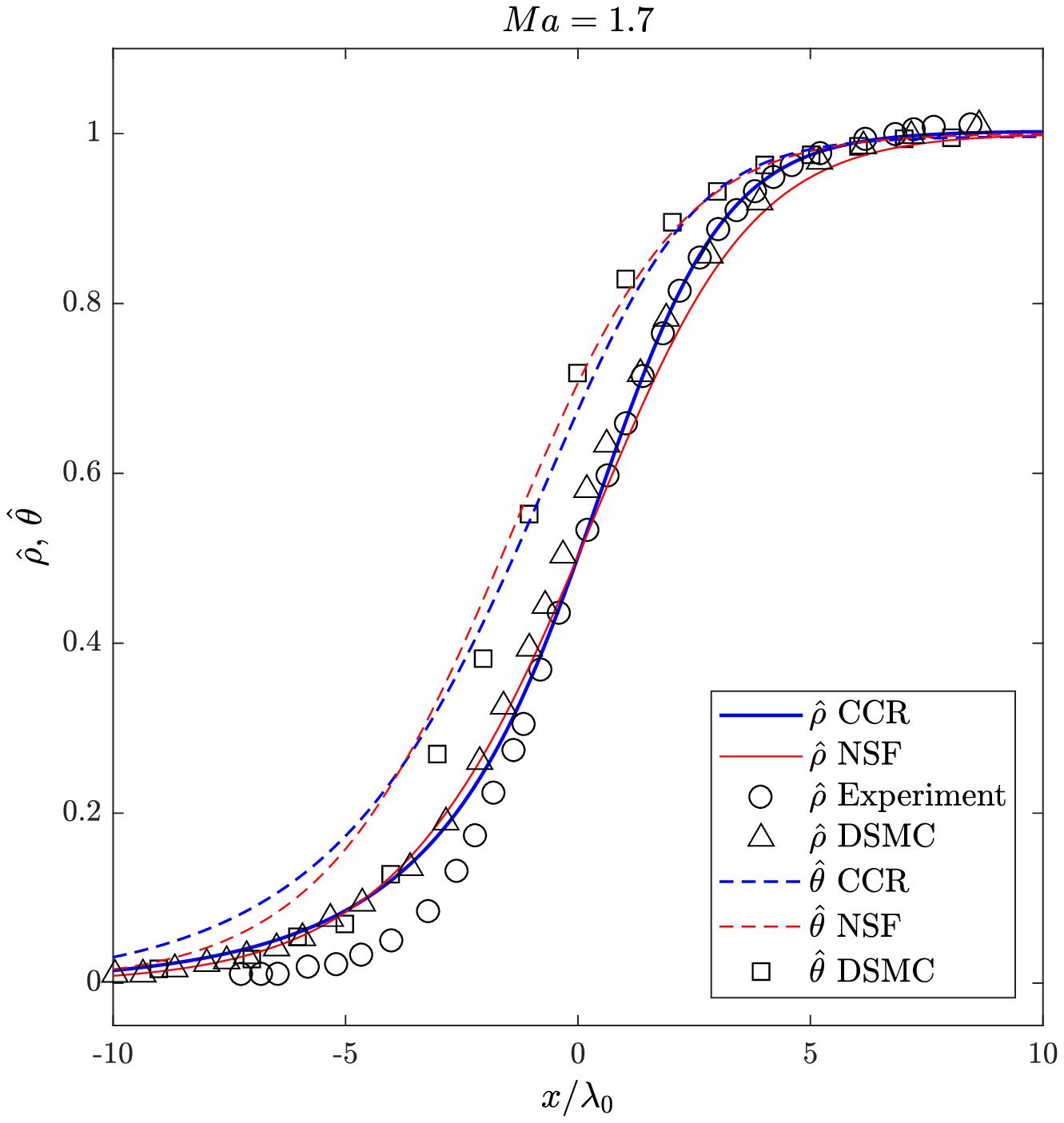}
  \caption{}
  \label{fig:sub1}
\end{subfigure}%
\begin{subfigure}{.33\textwidth}
  \centering
  \includegraphics[width=\linewidth]{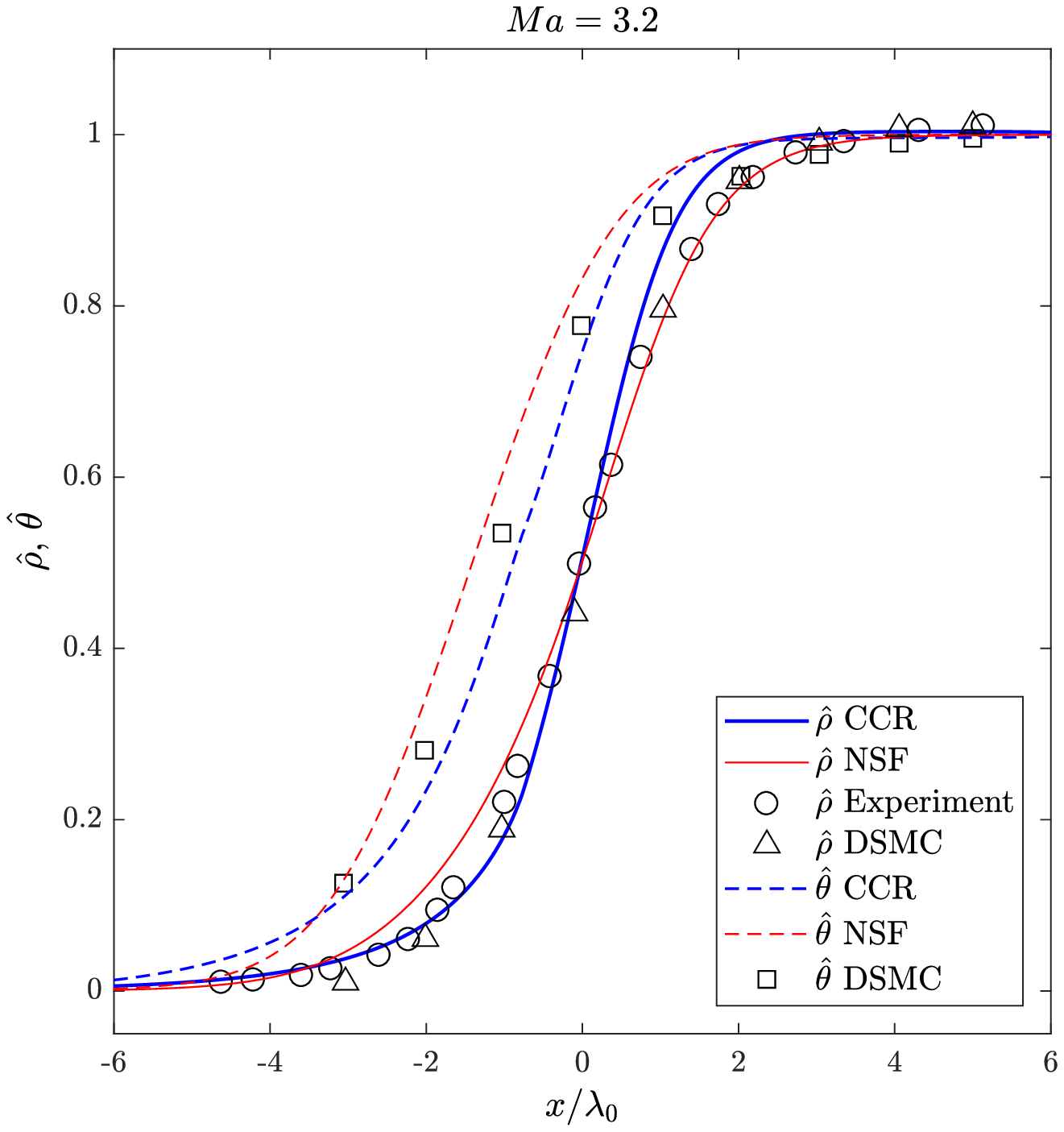}
  \caption{}
  \label{fig:sub2}
\end{subfigure}
\begin{subfigure}{.33\textwidth}
  \centering
  \includegraphics[width=\linewidth]{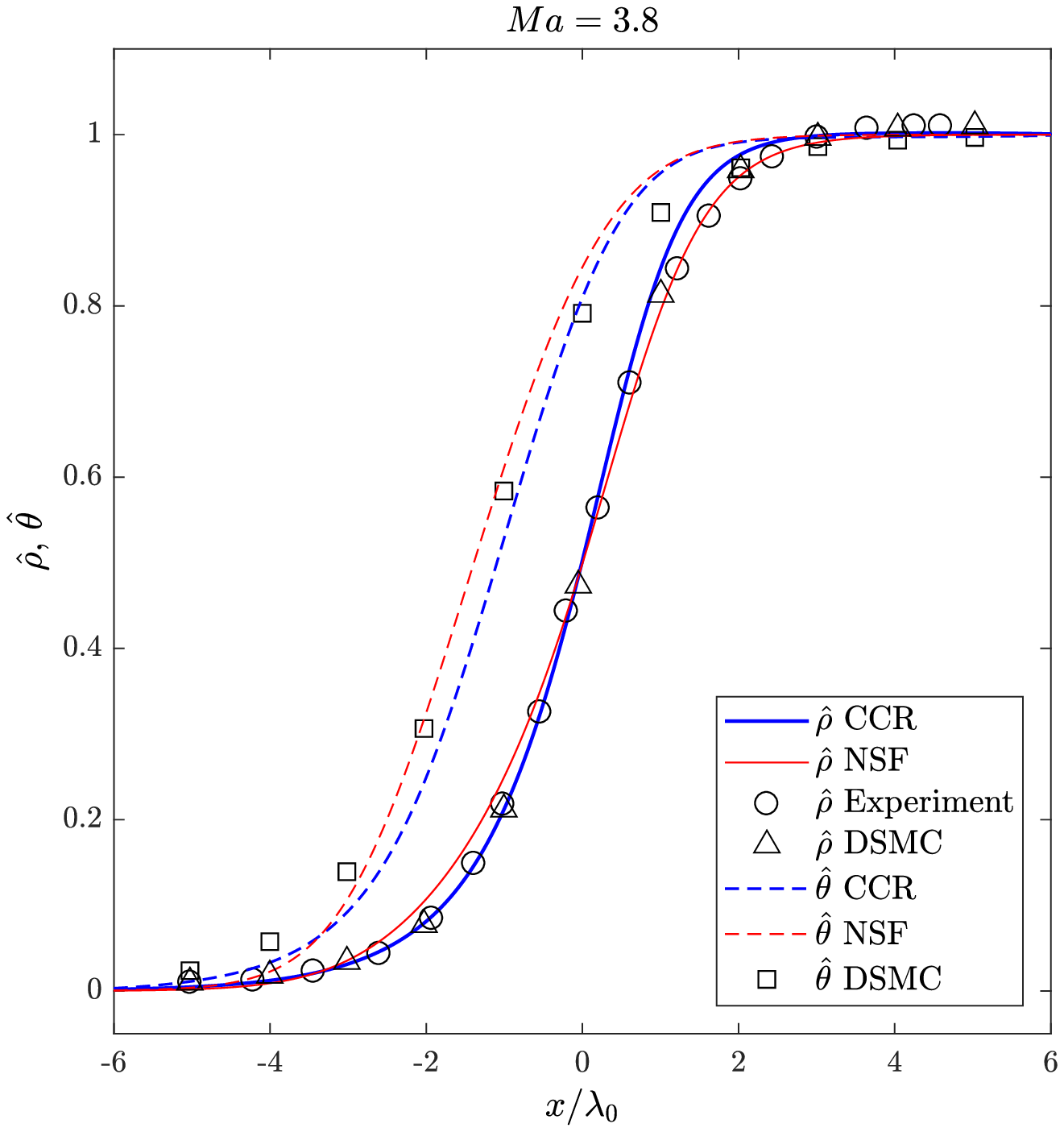}
  \caption{}
  \label{fig:sub3}
\end{subfigure}
\caption{Comparison of reduced density and temperature  profiles obtained using the polyatomic-CCR
models with the experimental measurements of \citet{alsmeyer1976} for Mach numbers: (a) $\mathrm{Ma} = 1.7$ (b) $\mathrm{Ma} = 3.2$ and (c) $\mathrm{Ma} = 3.8$.}
\label{fig:shock}
\end{figure}

Figure (\ref{fig:shock}) shows the profiles of reduced density, $\hat{\rho}:=(\rho-\rho_0)/(\rho_\infty-\rho_0)$ and reduced temperature,  $\hat{\theta}:=(\theta-\theta_0)/(\theta_\infty-\theta_0)$ vs $x/\lambda_0$ for three Mach
numbers, $\mathrm{Ma} = 1.7$ (\ref{fig:shock}.a), $\mathrm{Ma} =3.2$ (\ref{fig:shock}.b) and $\mathrm{Ma} =3.8$ (\ref{fig:shock}.c). For a smaller Mach number ($\mathrm{Ma} = 1.7$), the macroscopic theories and the DSMC give a good agreement.  At higher Mach numbers (\ref{fig:shock}.b, c), the upstream region of the shock is
resolved accurately by the polyatomic-CCR model whereas some deviations are observed in the downstream region. Interestingly, for the NSF theory we observe an opposite behaviour. For the temperature shock profiles, the DSMC predictions appear to be slight narrower than the macroscopic theories. The NSF theory slightly over-predicts the temperature while the polyatomic-CCR theory under-predicts it. 
The main goal of the present study was to derive the polyatomic-CCR model and demonstrate its usefulness for some practical applications; a thorough  test of these equations for the shock wave flow problems is beyond the scope of this article. Nevertheless, for the considered range of the Mach number, the overall results from the polyatomic-CCR model are in good agreement with the experimental data. 

This is noteworthy given that the entropy current density we assume has been derived in  \citep{pavic2013maximum}. It employs the exponential distribution function in kinetic theory and the entropy of this distribution is analytically obtained by integrating over internal degrees of freedom and particular velocity. This procedure crucially depends on the concept of local temperature which has been used in the exponential distribution. Despite the difficulties in this concept, it turns out that local temperature may thus be fruitfully applied to rarefied flows of poly-atomic gases as well.
% \begin{figure}
% \centering
% \begin{subfigure}{.5\textwidth}
%   \centering
%   \includegraphics[width=\linewidth]{mach1p7.eps}
%   \caption{}
%   \label{fig:sub1}
% \end{subfigure}%
% \begin{subfigure}{.5\textwidth}
%   \centering
%   \includegraphics[width=\linewidth]{mach3p2.eps}
%   \caption{}
%   \label{fig:sub2}
% \end{subfigure}
% \caption{Comparison of reduced density and temperature  profiles obtained using the polyatomic-CCR
% models with the experimental measurements of Alsmeyer (1976) for two different Mach numbers: (a) $\mathrm{Ma} = 1.7$ (b) $\mathrm{Ma} = 3.2$.}
% \label{fig:shock}
% \end{figure}

% \begin{figure}
% \centering
% \begin{subfigure}{.5\textwidth}
%   \centering
%   \includegraphics[width=\linewidth]{mach3p8.eps}
%   \caption{A subfigure}
%   \label{fig:sub1}
% \end{subfigure}%
% \begin{subfigure}{.5\textwidth}
%   \centering
%   \includegraphics[width=\linewidth]{mach3p8.eps}
%   \caption{A subfigure}
%   \label{fig:sub2}
% \end{subfigure}
% \caption{A figure with two subfigures}
% \label{fig:test}
% \end{figure}

\section{Conclusion and future directions}
%It is indeed appealing to take analytical approach developed for the classical fluid dynamics, but there is no need to sacrifice a lot of accuracy just to enforce them.
% , since if one can model the process more accurately while maintaining the same assumptions and thermodynamic structure.
CIT framework applied to rarefied flows require justification for the local equilibrium hypothesis which is taken to hold in the case of rarefied gases. There have been several objections to that assumption, especially regarding the concept of local temperature and entropy. However, under the RET assumption of entropy current density subject to a different constitutive law, CCR modelling of rarefied gas flows extends the applicability beyond classical NSF equation whilst maintaining the same mathematical and thermodynamical structure. Good agreement of CCR has been previously obtained with experiments in the case of monoatomic gases.
In this article, we have proposed a polyatomic-CCR model,  and obtained its solutions for some classical flow problems, namely slow flow past an aerosol of different shapes. Further these have been benchmarked with experimental data from the literature. The polyatomic-CCR theory provides an excellent match with experimental observations over the rage of all Knudsen numbers. 
We have also computed the normal shock structure using the polyatomic-CCR model. A very good agreement was observed with the kinetic theory, as well as with the experimental data. 

We have formulated Green's functions and a MFS technique for the polyatomic-CCR model, which  in principle, allow for simulations of slow polyatomic gas flow over any arbitrary shapes and sizes. Thus, from a modelling perspective, a simultaneously computationally efficient and accurate (to nano-scales) method down was proposed. For a future direction, a machine learning algorithm trained using the data from the polyatomic-CCR framework is planned, which can provide simulation-for-design tool to study the multi-scale process of dispersion and dynamics of UFPs in air.

 The proposed model is found to possess three important limitations which could be worked upon in the future. They involve thermodynamic issues related to local equilibrium, temperature dependent specific heats and asymptotic consistency of higher order boundary conditions. For example, the last issue assumes importance when one deals with flows with thermal transpiration (eg., Knudsen pump). 
 We comment upon these below. 
The proposed model assumes the validity of local equilibrium.
This assumption cannot be justified when the bulk-to-shear-viscosity ratio is very large; for example, the carbon dioxide gas has the bulk-to-shear-viscosity ratio $\sim 2000$. Conceptually, in addition, the definition of local thermodynamic quantities has been questioned. Temperature and entropy have been emphasized the most in this regard. The proportionality of internal energy and temperature has been especially a crucial assumption which requires to be addressed.  In a scenario like this, an interesting approach could be to use an extended Gibbs' equation  in order to get a refined NSF equations or the corresponding coupled constitutive relations \citep{Ruggeri2015NONLINEARME}.

The polyatomic-CCR theory developed here assumes the constant specific heats (hence $\delta$ to be constant).  In hindsight of the results in the Section \ref{sec: normal shock structure in nitrogen}, it is worthwhile to discuss the validity of this assumption. For the range of Mach number considered in this article ($1.7 \le \mathrm{Ma} \le 4$) the maximum temperature variations---from (\ref{Rankine-Hugoniot relations})---are $1.46 \le \frac{\theta_{\infty}}{\theta_0} \le 4.0s$, which are seemingly high. However, for the nitrogen gas, the change in $\delta$ between 200K--800K is about $\lesssim 7.5\%$ \citep{borgnakke2009fundamentals}, therefore the assumption of $\delta$ being constant was justifiable. Nevertheless, for large enough temperature variations, this assumption may not hold true and an extension of the CCR model for temperature dependent specific heats (i.e., assuming $\delta = \delta(\theta)$) will be sought. 

Furthermore, the boundary conditions (\ref{2nd boundary condition}) employed in Section \ref{sec: Slow rarefied gas flow}-\ref{sec: Slow rarefied gas flow doublet} were obtained in a way to fit the drag coefficient on the sphere (by choosing appropriate value of the second-order slip coefficient). The asymptotic analysis (see also, Appendix \ref{sec: Asymptotic analysis of the boundary conditions}) suggests that these boundary conditions may lead to erroneous predictions in situations when thermal transpiration/thermal-stress play important role. A detailed analysis of such flows along with a consistent set of boundary conditions for the polyatomic-CCR theory is also planned for the future.

%  This assumption can be justified under the condition: $\max(\tau_T,\tau_R)\ll \tau_P$, where $\tau_t$, $\tau_t$,  (i) the ratio of translational relaxation time is sufficiently smaller than a microscopic time, and  (ii) the ratio of 
%\bibliographystyle{jfm}
%\bibliography{jfm}
%Use of the above commands will create a bibliography using the .bib file. Shown below is a bibliography built from individual items.

% \bibliographystyle{jfm}
% %\bibliography{jfm2esam}
\section{Funding}
This work has been financially supported by ``SRG" project SRG/2021/000790 funded by the Science \& Engineering Research Board, India.

\section{Declaration of interests} The authors report no conflict of interest.

\appendix

\section{Asymptotic analysis of the boundary conditions}
\label{sec: Asymptotic analysis of the boundary conditions}
Here, we shall carry out a comparison of boundary conditions (\ref{2nd boundary condition}) with
those from the literature and discuss qualitative effects of different terms
appearing in the boundary conditions. The (linearized and dimensionless) generalized slip boundary condition
relates the tangential gas velocity slip $\hat{V}_{i}$ to the tangential
shear stress and heat flux, as \citep{Lockerby2004}%
\begin{eqnarray}
\hat{V}_{i}t_{i} &=&A_{1}\sqrt{\frac{\pi }{2}}\mathrm{Kn}\left( \frac{\partial \hat{v}%
_{i}}{\partial \hat{x}_{j}}+\frac{\partial \hat{v}_{j}}{\partial \hat{x}_{i}}%
\right) n_{i}t_{j}-A_{2}\left( \sqrt{\frac{\pi }{2}}\mathrm{Kn}\right) ^{2}\frac{%
\partial }{\partial \hat{x}_{k}}\left( \frac{\partial \hat{v}_{i}}{\partial 
\hat{x}_{k}}+\frac{\partial \hat{v}_{k}}{\partial \hat{x}_{i}}\right) t_{i}\notag
\\
&&+A_{3}\mathrm{Kn}\frac{\partial \hat{\theta}}{\partial \hat{x}_{i}}t_{i}-A_{4}\sqrt{%
\frac{\pi }{2}}\mathrm{Kn}^{2}\frac{\partial ^{2}\hat{\theta}}{\partial \hat{x}_{i}%
\hat{x}_{j}}n_{i}t_{j}\text{,}
\label{generalized slip boundary condition}
\end{eqnarray}%
where $n_{i}$ is unit normal pointing from the boundary into the gas and $%
t_{i}$ is a tangential direction. In the last equation, $A_{1}$ and $A_{2}$ are
the first-order and second-order slip coefficients, respectively. Furthermore, $A_{3}$ is
thermal slip coefficient \citep{SharipovData} \footnote[1]{In Sharipov's  notation $A_{3}=\sigma _{T}$.},
which gives rise to thermal creep, and the last term (with $A_{4}$) is due
to the thermal-stress slip flow \citep{Lockerby2004}. \footnote[2]{In Lockerby's
notation the mean free path is defined as $\lambda =\sqrt{\pi /2}\mathrm{Kn}L$, and in
Sharipov's notation $A_{1}=2\sigma _{P}/\sqrt{\pi }$ where $\sigma _{P}$ is
defined as the viscous slip coefficient in \cite{SharipovData} Table 1, and $\ell$
is defined by as $\ell=\sqrt{2}\mathrm{Kn}L$.}

The boundary conditions (\ref{2nd boundary condition}), when written in the Cartesian coordinates, read 
\begin{eqnarray}
\hat{\sigma}_{ij}n_{i}t_{j} &=&-\sqrt{\frac{2}{\pi }}\frac{\chi }{2-\chi }%
\frac{1}{\eta _{\mathrm{VS}}}\left( \hat{V}_{i}+\eta ^{1}\hat{q}_{i}\right)
t_{i}\text{,} \label{BC1A}\\
\hat{q}_{k}n_{k} &=&-\sqrt{\frac{2}{\pi }}\frac{\chi }{2-\chi }\frac{%
4+\delta }{2}\frac{1}{\eta _{\mathrm{TJ}}}\left( \mathcal{\hat{T}}+\eta ^{1}%
\hat{\sigma}_{ij}n_{i}n_{j}\right)\text{.} \label{BC2A} 
\end{eqnarray}%
Here, we have further introduced the accommodation coefficient $\chi $, the
velocity-slip coefficient $\eta _{\mathrm{VS}}$ and the temperature-jump
coefficient $\eta _{\mathrm{TJ}}$. For the calculations
performed in Section \ref{sec: Slow rarefied gas flow}-\ref{sec: Slow rarefied gas flow doublet}, we had taken $\chi =\eta _{\mathrm{VS}}=\eta _{%
\mathrm{TJ}}=1$ and $\eta ^{1}=3/4$; below we provide a rationale for this choice.

Substituting the CCR (\ref{Linear constitutive}) into (\ref{BC1A})
 and using the conservation laws (\ref{Linear conservation laws}), one gets%
\begin{eqnarray}
\hat{V}_{i}t_{i} &=&\sqrt{\frac{\pi }{2}}Kn\frac{2-\chi }{\chi }\eta _{%
\mathrm{VS}}\left( \frac{\partial \hat{v}_{i}}{\partial \hat{x}_{j}}+\frac{%
\partial \hat{v}_{j}}{\partial \hat{x}_{i}}\right) n_{i}t_{j}+\eta ^{1}\frac{%
Kn}{\Pr }\frac{\partial \hat{\sigma}_{ik}}{\partial \hat{x}_{k}}t_{i} \notag\\
&&+\frac{\eta ^{1}c_{p}}{\Pr }Kn\frac{\partial \hat{\theta}}{\partial \hat{x}%
_{i}}t_{i}+\sqrt{\frac{\pi }{2}}\frac{2-\chi }{\chi }\eta _{\mathrm{VS}}Kn%
\frac{2}{5+\delta }\left( \frac{\partial \hat{q}_{i}}{\partial \hat{x}_{j}}+%
\frac{\partial \hat{q}_{j}}{\partial \hat{x}_{i}}\right) n_{i}t_{j}\text{.}
\end{eqnarray}
In order to compare the last equation with (\ref{generalized slip boundary condition}), $\hat{\sigma}_{ik}$ and $%
\hat{q}_{i}$ need be replaced by the corresponding NSF equations, i.e., by $%
\hat{\sigma}_{ik}^{\mathrm{NSF}}$ \ and $\hat{q}_{i}^{\mathrm{NSF}}$, to get 
\begin{eqnarray}
\hat{V}_{i}t_{i} &=&\sqrt{\frac{\pi }{2}}Kn\frac{2-\chi }{\chi }\eta _{%
\mathrm{VS}}\left( \frac{\partial \hat{v}_{i}}{\partial \hat{x}_{j}}+\frac{%
\partial \hat{v}_{j}}{\partial \hat{x}_{i}}\right) n_{i}t_{j}-\eta ^{1}\frac{%
Kn^{2}}{\Pr }\frac{\partial }{\partial \hat{x}_{k}}\left( \frac{\partial 
\hat{v}_{i}}{\partial \hat{x}_{k}}+\frac{\partial \hat{v}_{k}}{\partial \hat{%
x}_{i}}\right) t_{i} \notag\\
&&+\frac{\eta ^{1}c_{p}}{\Pr }Kn\frac{\partial \hat{\theta}}{\partial \hat{x}%
_{i}}t_{i}+\sqrt{\frac{\pi }{2}}Kn\frac{2-\chi }{\chi }\eta _{\mathrm{VS}}%
\frac{2}{5+\delta }\left( \frac{\partial \hat{q}^{\mathrm{NSF}}_{i}}{\partial \hat{x}_{j}}+%
\frac{\partial \hat{q}^{\mathrm{NSF}}_{j}}{\partial \hat{x}_{i}}\right) n_{i}t_{j}\text{.}
\end{eqnarray}%
Comparing the last equation with a generalized slip boundary condition (\ref{generalized slip boundary condition}),
one identifies%

\begin{equation}
    A_{1}=\frac{2-\chi }{\chi }\eta _{\mathrm{VS}}\text{, }A_{2}=\frac{2}{\pi }%
\frac{\eta ^{1}}{\Pr }\text{, and }A_{3}=\frac{\eta ^{1}c_{p}}{\Pr }=\sigma _{T}\text{.}
\end{equation}

\begin{table}
  \begin{center}
\def~{\hphantom{0}}
  \begin{tabular}{lccl}
         & $A_{1}$   &   $A_{2}$ & Model \\[3pt]
\citet{DEISSLER1964681} & $1$ & $9/8$ & Monoatomic/polyatomic \\ 
\citet{StruchtrupandTorrilhon2008} & $1$ & $0.531$ & Monoatomic/R13 \\ 
\citet{LBM2011} & $1$ & $0.8$ & Lattice Boltzmann \\ 
\citet{Gibelli2012} & $1.1144$ & $0.4952$ & Monoatomic/hard-sphere Boltzmann \\ 
\citet{Lockerby2004} & $1$ & $0.145-0.19$ & Monoatomic/polyatomic \\ 
\citet{Hadjiconstantinou2003} & $1.11$ & $0.61$ & Monoatomic/hard-sphere Boltzmann \\ 
\citet{SharipovData} & $1.1-1.14$ & -- & BGK/hard-sphere Boltzmann \\ 
Present model & $1$ & $0.682$ & CCR theory%
  \end{tabular}
  \caption{Values of the first-order and
second-order slip coefficients \citep{reviewslip2012}.}
  \label{tab:slip coefficients}
  \end{center}
\end{table}

There exits a vast literature documenting values of the first-order and
second-order slip coefficients \citep{reviewslip2012}; few are tabulated in Table \ref{tab:slip coefficients}. As we can see from the table, the value for $A_{1}$
is around unity, however, no conclusion can be reached on the correct value
of the coefficient $A_{2}$; the value for $A_{2}(=\frac{3}{2\pi }\frac{1}{%
\Pr })$ was chosen so that it gives good match for the drag coefficient for sphere. However, for the chosen values of $A_{2}$, the presented boundary conditions yields  
$A_{3}=\frac{3}{4}\frac{c_{p}}{\Pr }$, which is about 3.5 times as then the
reported values; in \citep{SharipovData}, the value $A_{3}$ ranges between $0.94-1.12$.
Therefore, one would expect that the presented model/boundary conditions
will not provide adequate results for problems involving thermal
creep/thermal-stress. In order to tackle such problems, one might choose $%
\eta ^{1}~$appropriately (instead of $3/4$) to get the correct value for $%
A_{3}$; however, this is beyond the scope of this article.

Similarly, substituting the CCR (\ref{Linear constitutive}) into (\ref{BC2A}) and using the
conservation laws, one obtains%
\begin{equation}
    \mathcal{\hat{T}}=\zeta _{T}\sqrt{2}Kn\frac{\partial \hat{\theta}}{\partial 
\hat{x}_{k}}n_{k}+\sqrt{\frac{\pi }{2}}\frac{2-\chi }{\chi }\frac{2\eta _{%
\mathrm{TJ}}}{4+\delta }\frac{Kn}{\Pr }\frac{\partial \hat{\sigma}_{kr}}{%
\partial \hat{x}_{r}}n_{k}-\eta ^{1}\hat{\sigma}_{ij}n_{i}n_{j}\text{,}
\end{equation}
where 
\begin{equation}
    \zeta _{T}=\frac{\sqrt{\pi }}{2}\frac{2-\chi }{\chi }\frac{2\eta _{\mathrm{TJ%
}}}{4+\delta }\frac{c_{p}}{\Pr }\text{.}
\end{equation}
Here, $\zeta _{T}$ is the temperature jump coefficient \citep{SharipovData}. The
reported value for $\zeta _{T}$ by \citet{SharipovData} is $\zeta _{T}=\frac{\sqrt{\pi }}{2}%
\frac{2}{4+\delta }\frac{c_{p}}{\Pr }\left( \frac{2-\chi }{\chi }%
+0.17\right) $; which is about $17\%$ higher than considered value. 

\begin{figure}
\centering
\begin{subfigure}{.55\textwidth}
  \centering
  \includegraphics[width=\linewidth]{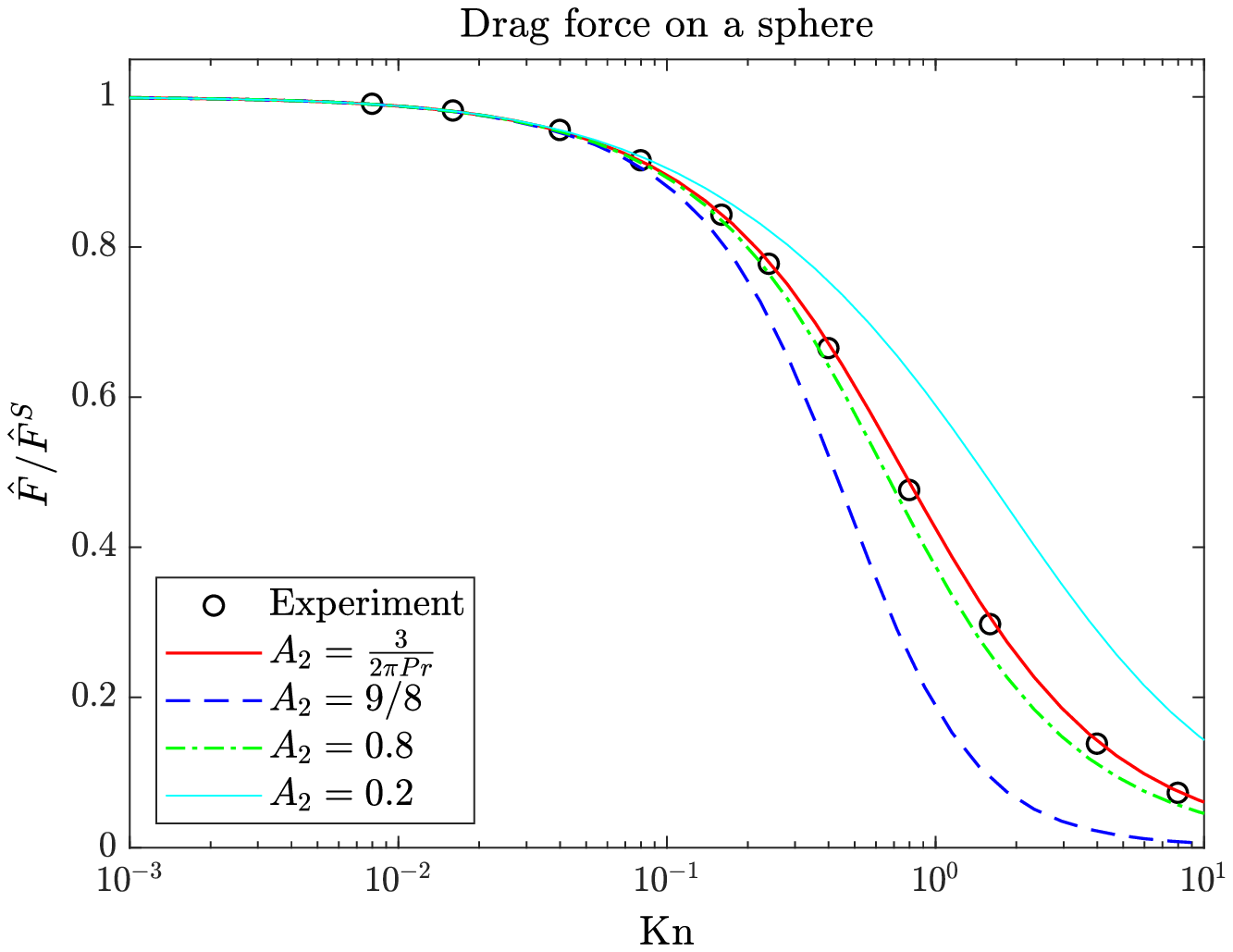}
  \caption{The normalised drag force  on a sphere  against Knudsen number. The results predicted by the CCR theory with different values of $A_2$, while taking $\eta_{\mathrm{VS}}=\eta_{\mathrm{TJ}}=1$.}
  \label{fig:dragsphereA2}
\end{subfigure}%
\begin{subfigure}{.55\textwidth}
  \centering
  \includegraphics[width=\linewidth]{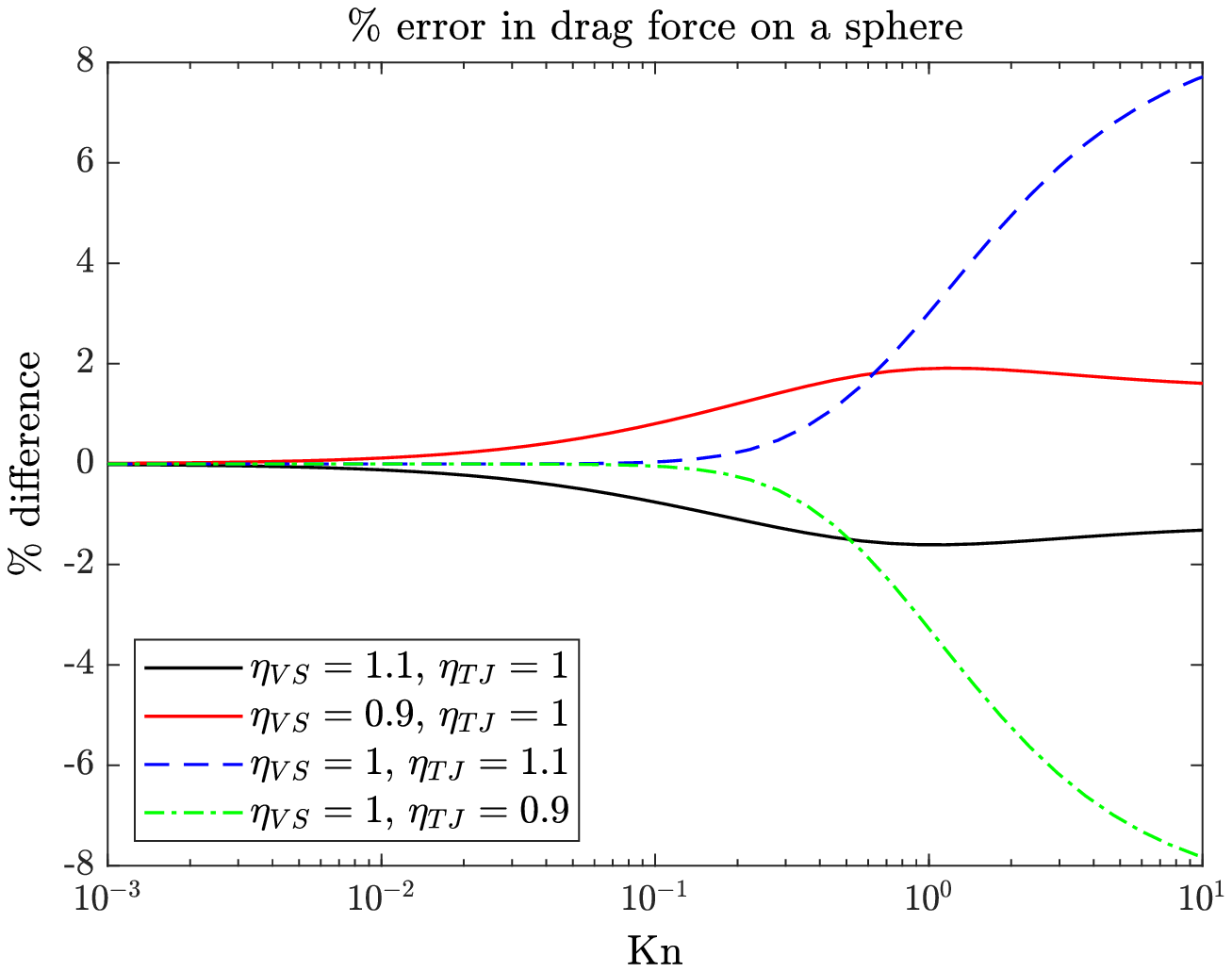}
  \caption{Relative percentage error in drag force predicted by the CCR theory for $A_{2}=\frac{3}{2\pi }\frac{1}{%
\Pr }$ for different values of $\eta_{\mathrm{VS}}$ and $\eta_{\mathrm{TJ}}$.}
  \label{fig:dragsphereslipjump}
\end{subfigure}%
\caption{}
\label{fig:test}
\end{figure}

In order to quantify the change in drag coefficient with respect to the slight variations in  $A_2$ (the second-order slip coefficient), $\eta_{\mathrm{VS}}$, $\eta_{\mathrm{TJ}}$, in Fig.~(\ref{fig:dragsphereA2}) we plot the drag on a sphere with different values of $A_2$, while taking $\eta_{\mathrm{VS}}=\eta_{\mathrm{TJ}}=1$. The second-order slip coefficient has a notable effect on the drag, in particular at large values for the Knudsen number. On the other hand, the small variations in values for $\eta_{\mathrm{VS}}$ $\eta_{\mathrm{TJ}}$ lead to slight change in drag coefficients---as seen in Fig.~(\ref{fig:dragsphereslipjump})---which is less than 10\% within the reported values of $\eta_{\mathrm{VS}}$ and $\eta_{\mathrm{TJ}}$.

% \begin{thebibliography}{99}
\bibliography{jfm.bib}
\bibliographystyle{jfm}
% \end{thebibliography}

%% End of file `jfm2esam.bib'.

\end{document}